\documentclass[iop,apj,numberedappendix,appendixfloats,twocolappendix]{emulateapj}
\usepackage{apjfonts}
\usepackage{amsmath}
\usepackage{iondefs}
\usepackage{symbols}

\shorttitle{\sc Ionization Modeling}
\shortauthors{\sc Churchill {\etal}}

\begin{document}

\title{Ionization Modeling Astrophysical Gaseous Structures. I.
  The Optically Thin Regime}

\author{\sc
Christopher W. Churchill\altaffilmark{1},
Elizabeth Klimek\altaffilmark{1}, 
Amber Medina\altaffilmark{1}, 
and
Jacob R. Vander Vliet\altaffilmark{1},
}

\altaffiltext{1}{New Mexico State University, MSC 4500, Las Cruces, NM 88003,
USA}


\begin{abstract}
We present a code for modeling the ionization conditions of optically
thin astrophysical gas structures.  Given the gas hydrogen density,
equilibrium temperature, elemental abundances, and the ionizing
spectrum, the code solves the equilibrium ionization fractions and
number densities for all ions from hydrogen to zinc.  The included
processes are photoionization, Auger ionization, direct collisional
ionization, excitation auto-ionization, charge exchange ionization,
two-body radiative recombination, dielectronic recombination, and
charge exchange recombination.  The ionizing spectrum can be
generalized to include the ultraviolet background (UVB) and/or
Starburst99 stellar populations of various masses, ages,
metallicities, and distances.  The ultimate goal with the code is to
provide fast computation of the ionization conditions of gas in N-body
+ hydrodynamics cosmological simulations, in particular adaptive mesh
refinement codes, in order to facilitate absorption line analysis of
the simulated gas for comparison with observations.  To this end, we
developed a method to linearize the rate equations and efficiently
solve the rate matrix with a minimum number of iterations.  Comparing
the code to Cloudy 13.03 (Ferland), we find that the hydrogen and
helium ionization fractions and the metal species ionization
corrections are in excellent agreement.  We discuss the science
drivers and plans for further development of the ionization code to a
full radiative hydrodynamic routine that can be employed for
processing the simulations.  A stand-alone version of the code has
been made publicly available.
\end{abstract}


\section{Introduction}
\label{sec:intro}

The evolution of galaxies is intimately linked to their gas processes.
Star formation rates are sustained by accretion of infalling gas
\citep[e.g.,][]{review14} and regulated by stellar feedback processes
\citep[e.g.,][]{stinson07, ceverino09, ceverino13, hopkins13a,
hopkins13b, kim13, st13-dwarfs}.  Accretion and feedback likely
persist in a stochastic quasi-balancing act that regulates galaxy
evolution and yields the global properties of galaxies
\citep[e.g.,][]{dave11a, dave11b, lilly13}, such as the stellar
mass-halo mass relation and the average star formation history, both a
function of halo mass and redshift \citep[e.g.,][]{behroozi10,
behroozi13, moster13}, the stellar mass-metallicity relation
\citep[e.g.,][]{tremonti04, mannucci10, bothwell13, gonzalez14}, and
the distribution of galaxies on the color-stellar mass diagram
\citep[e.g.,][]{red-herring14}.

With deeper appreciation for the key role gas plays in governing the
evolution of galaxies, concentrated effort has been focused on
incorporating increasingly sophisticated treatment of hydrodynamics
and stellar formation and feedback processes in cosmological
simulations.  The ultimate goal is to form realistic galaxies having
properties consistent with global galaxy relations while gaining
insight into the physics that yields these relations
\citep[e.g.,][]{ceverino09, ceverino13, hopkins13a, munshi13,
st13-dwarfs, agertz14}.

Since the flow of gas in and around galaxies is central to regulating
the evolution of galaxies, it is imperative the properties of this
``circumgalactic medium'' (CGM) associated with simulated galaxies
also match observations \citep[e.g.,][]{ford13a, ford13b, hummels13,
cwc-qals14}.


The CGM is observed using absorption lines in the spectra of luminous
background sources, such as quasars, whose lines of sight
serendipitously probe the vicinity of foreground galaxies.  Commonly
observed absorption lines include {\HI} {\Lya}~$\lambda 1215$
\citep[e.g.,][]{lanzetta95, stocke13, tumlinson13, mathes14}, the
{\CIVdblt} and {\OVIdblt} doublets \citep[e.g.,][]{simcoe06, stocke06,
fox07-ovi, fox07-civ, tumlinson11, stocke13, mathes14}, and the
{\MgIIdblt} doublet \citep[e.g.,][and references
therein]{nielsen-cat1, nielsen-cat2}.  Commonly measured absorption
line quantities include the equivalent widths, column densities, the
velocity spreads, the line-of-sight velocities with respect to the
presumed host galaxy, and, if Voigt profile (VP) decomposition is
undertaken, the number of VP components and their column densities,
velocities, and Doppler $b$ parameters (line profile broadening and
indicator of gas temperature and turbulence).  On a sightline by 
sightline basis, these quantities can be examined as a function 
of host galaxy properties (masses, luminosities, colors, star 
formation rates, etc.) and projected distance from the host 
galaxies.

Since the absorption occurs from ionized atoms and is proportional to
the product of the ion density and line of sight pathlength through
the gas, ionization modeling is required to infer the hydrogen density
and metallicity of the CGM absorbing gas from observed spectra.
The most commonly employed ionization code is Cloudy \citep{ferland98,
ferland13}.  Assuming the spectral energy distribution (SED) of the
ionizing spectrum is known, and assuming a cloud geometry, Cloudy
allows the ionization conditions, metallicity, and hydrogen density to
be constrained from the measured column densities. Usually, a
\citet{haardt11} ultraviolet background (UVB) ionizing spectrum is
assumed for the model clouds.

Ionization modeling is equally critical for the simulations.  Since
analysis of the simulated CGM requires a comparison with observed
absorption line properties, we are required to generate synthetic
absorption profiles of ``sightlines'' through the simulated CGM.  This
requires that the ionization fractions of the absorbing ions are
known.  There are two approaches.  The first is post-process
ionization modeling of the simulation output.  The second is
incorporation of the ionization modeling self-consistently into the
hydrodynamics of the cosmological simulation, which is computationally
expensive.

For both observations and simulations, several assumptions are often
employed, most importantly that the gas is in ionization equilibrium
and that the SED of the ionizing spectrum is known, at least
approximately.  In the CGM, the ionizing SED can have substantial
variations, both temporally and spatially.  Since the ionization
conditions strongly depend on the density and temperature of the
simulated gas, it is important the heating and cooling rates are as
realistic as possible.  However, the heating and cooling rates depend
upon the ionizing SED, the gas density, temperature, and the atomic
abundances of metals in the gas.  If the cooling time is shorter than
the ionization and/or recombination timescales of a given ion, then
that ion cannot settle into ionization equilibrium.  These
considerations highlight the importance for incorporating a spatially
resolved radiative transfer (RT) scheme that is coupled to the
hydrodynamics in cosmological simulations \citep[hereafter, radiation
  hydrodynamics, RH, see][]{mihalas2-book}.

Various explorations, often focused on specific astrophysical
problems, have been researched to examine the effects of implementing
RT and RH in hydrodynamic simulations \citep{howell03, iliev06,
  iliev09, whalen06, aubert08, finlator09, norman09, reynolds09}.  In
fact, progress has been made toward incorporating RH into some
cosmological simulations, either by solving the optically thin
variable Eddington tensor \citep{gnedin01, razoumov06, petkova09,
  rosdahl13}, or by employing ray-tracing methods \citep{abel02,
  wise11}.  Typically, only hydrogen and helium ionization balance is
treated.  However, \citet{cen06} studied the evolution of {\OIV} to
{\OIX} by directly integrating the rate equations using RH in their
simulations.  \citet{oppenheimer13} implemented RH into their
simulations and incorporated a rate matrix including several important
metals species, but so far they have limited their analysis to a
single parcel of gas independent of hydrodynamics.





To date, studies of the simulated CGM that employ the absorption line
technique have implemented the post-processing approach
\citep[e.g.,][]{ford13a, ford13b, hummels13, cwc-qals14}.  Typically,
the ionization code Cloudy is used to create a grid of ``cloud''
models as a function of redshift assuming the \citet{haardt11} UVB
ionizing SED for a range of gas-state variables. The cloud models
usually are constant density with a plane parallel geometry, are
illuminated by the ionizing SED on one face, and omit dust and cosmic
ray heating.  The ionization condition of the cloud model is normally
specified by the ionization parameter,
\begin{equation}
U = \frac{n_{\gamma}}{n\subH} = \frac{4\pi}{n\subH} 
\int _{\nu_0}^{\infty} \frac{J_{\nu}}{h\nu} \, d\nu \, ,
\label{eq:ionpar}
\end{equation}
which is defined at the illuminated edge of the cloud model, where
$n\subH$ is the constant hydrogen density, $J_\nu$ is specific
intensity of the ionizing SED, and $h\nu_0$ is the ionization
threshold energy for neutral hydrogen.  Since Cloudy solves the RT
through a multi-zone cloud model, the ionization structure can vary
with depth into the cloud.  The standard output of Cloudy are the
zone-depth weighted {\it average\/} ionization fractions.  A fixed
metallicity and abundance pattern (usually solar) is adopted.  For a
given parcel of gas in the simulations, the Cloudy grid serves as a
look-up table for ionization fractions, from which the metallicity and
abundance pattern can be scaled \citep[e.g.,][]{bs86, dittmann95} to
yield the column density of desired ions required for generating
synthetic absorption line spectra.

\citet{ford13a} employ a post-processing scheme in which they use
Cloudy 08.00 look-up tables to determine the metal species ionization
fractions as a function of density, temperature, and the UVB ionizing
spectrum.  For gas particles with large neutral hydrogen column
densities, i.e., $N(\HI) > 18$, they incorporate a correction factor
for self-shielding by matching the neutral hydrogen fractions to the
models of \citet{fagig09}.  They assume all of the column density from
the Mg$^{+}$ ion arises in the self-shielded region.
\citet{hummels13} track the hydrogen and helium ionization states
within their simulations, and then use a Cloudy 07.02.01 look-up table
to determine the post-processed ionization fraction of the metal ions
as a function of density, temperature, and UVB ionizing
spectrum. Whereas \citet{ford13a, ford13b} post-processed only their
line of sight quantities, \citet{hummels13} post-processed their full
simulation output and obtain the spatial distribution of the ionic
species.

Using a somewhat different approach, \citet{fumagalli11} adopted a
post-processing Monte-Carlo RT method that removes the ``preferred''
direction of the photon path inherent in Cloudy by accounting for
scattering.  Like Cloudy, their code includes collisional ionization
and photoionization, however they explore photoionization from both
the UVB and local stellar sources. They focus on neutral hydrogen
column densities and absorption profiles, though they did make rough
estimates of {\SiII} column densities and absorption strengths.  They
post-processed their full simulation output and explored the effects
of the various ionization mechanisms on the spatial distribution of
$N({\HI})$ in the vicinity of simulated galaxies by examining the
differences between collisional ionization only models, collisional
plus UVB photoionization models, and models that incorporate
collisional and UVB plus stellar photoionization.

\citet{cwc-qals14} applied an ionization code of their own design to
post-process the ionization conditions of their full simulation output
and conducted a pilot study of how well the inferred conditions of the
gas from simulated absorption lines compared with the actual
properties of the gas giving rise to the absorption.  Our goal with
this paper, is to present the details of the ionization code employed
by \citet{cwc-qals14}.  The code, called the hydroART Radiative
Algorithm for Trace Elements ({\code}), is designed to ultimately be
adaptable and implemented as post-processing RH for cosmological
simulations, especially the Eulerian N-body hydrodynamic code hydroART
\citep{kravtsovPhDT, kravtsov04}.  Currently, {\code} is applied as a
post-processing step to the full simulation output but does not yet
include RT nor RH through the simulation box.

In Section~\ref{sec:hydroART}, we describe the fundamental
characteristics of the hydroART code and our current approach to
post-process ionization modeling.  In Section~\ref{sec:themodel}, we
detail the physics incorporated into the {\code} and the method of
solution for the rate matrix.  We compare isolated {\code} cloud
models to those from the ``industry standard'' code Cloudy 13.03 in
Section~\ref{sec:compare}.  In Section~\ref{sec:conclude}, we provide
a general summary, including a description of planned future growth
and implementation of {\code}.  In addition to the first results from
an application of {\code} to the hydroART simulations presented in
\citet{cwc-qals14}, a stand alone version of the code has been
successfully applied to observed absorption line systems by
\citet{ggk-q1317} and \citet{cwc-q1317}.

\section{Hydrodynamic Simulations}
\label{sec:hydroART}

Underlying all hydrodynamic cosmological simulations is the
gravitating dark matter in an expanding simulation box, which is
reduced to an N-body problem that is commonly solved using
hierarchical multipole expansion (tree algorithms).  The baryons are
usually treated as an ideal fluid so that one can simplify the
hydrodynamics in terms of the Euler and continuity equations as
governed by the first law of thermodynamics.  Two main numerical
methods are employed to solve the coupled system of collisional
baryonic matter and collisionless dark matter: the particle methods,
which discretize mass, and grid-based methods, which discretize space.

A popular particle based method is smoothed particle hydrodynamics, or
SPH, which solves the Lagrangian form of the Euler equations.  Because
the gas is discretized into particles, SPH can achieve good spatial
resolutions in high-density regions, but is not as robust in
low-density regions. It also suffers from resolution degradation in
shock regions due to artificial viscosity \citep{agertz07}.

Alternatively, Eulerian grid-based methods solve the hydrodynamic
equations across a structured grid; each grid cell contains constant
properties (density, temperature, velocity, etc.) and represents a
finite volume of the gas fluid.  Grid-based methods are effective in
both high- and low-density regions and can handle shocks
\citep{dolag08}.  In the cosmological setting, adaptive mesh
refinement (AMR), in which grid cells vary in size in inverse
proportion to the gas density, are employed to increase the resolution
in regions of rapid evolution while optimizing the number of cells and
computational demands.

For our studies of the simulated CGM, we use the Eulerian N-body
hydrodynamic code hydroART \citep{kravtsovPhDT, kravtsov04}.  The
hydroART code follows the evolution of a gravitating N-body dark
matter halo and models the baryons using Eulerian hydrodynamics; it is
a grid-based AMR code that uses the zoom-in technique of
\citet{klypin01} and includes all of the currently relevant physics of
galaxy formation \citep{ceverino09, ceverino10, ceverino12,
  ceverino13, st13-dwarfs}.

Stars are formed deterministically with the observed low efficiency in
the cold and dense gas of molecular cloud environments. A stellar
particle represents a population of stars with a given mass, age, and
metallicity.  As the particles age, their mass decreases as supernovae
are converted back into gas.  The stellar feedback model includes the
major contributions from photoionization heating, direct radiation
pressure, energy from type II and type Ia supernovae, and stellar
winds \citep[see][for the most recent star formation and feedback
  recipes]{st13-dwarfs}.

The high-resolution region around the galaxy is typically $\sim 1$--2
Mpc across.  The hydrodynamics is resolved with $\simeq 7 \times 10^6$
grid cells, with a minimum cell size of roughly $30~h^{-1}$ pc at
$z=0$.  At each grid cell, the hydroART code follows the evolution of
the density, temperature, velocity, and metal mass fraction. The
metals produced in type II and Ia supernovae are followed separately
and are self-consistently advected with the gas flow.

The heating and cooling balance of the gas is determined using heating
and cooling functions obtained from Cloudy 8.00
\citep[see][]{ceverino13}.  These account for stellar ionizing SEDs
(as appropriate to the location and density of the grid cell and the
ages, masses, and metallicities of the stellar particles), ionization
by the UVB, molecular line cooling, and self-shielding of high column
density gas.  The treatment is similar to that implemented by
\cite{wss09}, but also includes the effects of ionization by stellar
radiation.


\subsection{Treatment of Ionization Balance}

We developed a code ({\code}) that performs equilibrium
ionization calculations in hydroART grid cells, with the goal of
incorporating it as a full RH treatment.  The main motivation for the
development of {\code} is to apply it to AMR cosmological simulation
in order to study the chemical and ionization conditions of the
circumgalactic medium in simulated galaxies using absorption line
techniques \citep[see][]{cwc-qals14}.  The code we present here
computes the equilibrium ionization fractions and number densities of
the ions in the gas, and is currently applied as a post-processing
step.

We treat each grid cell as an isolated ``cloudlette''.  The three
important gas properties associated with a cell are (1) the hydrogen
density, $n\subH$, (2) the equilibrium temperature, $T$, and (3) the
abundances of all atomic species.  In order to treat photoionization
processes, the spectral energy distribution (SED) of the ionizing
radiation must also be specified, which requires (4) the redshift, $z$,
which provides the cosmic epoch of the UVB ionizing radiation.  In
addition, the option to include stellar radiation is provided, which
requires (5) the characteristics (mass, age, metallicity, and
locations) of stellar populations.  When applied to hydroART, the
stellar populations are drawn from the stellar particles in the
simulated galaxy.

The present version of {\code} does not treat RT through the grid
cell, so there is no ionization structure within the grid cells (see
Section~\ref{sec:thingas}).  As such, currently, there is no
assumption about the gas geometry.  However, note that the commonly
employed ionization parameter (Eq.~\ref{eq:ionpar}) is well defined
and can be easily computed from the above inputs.

\section{The Ionization Model}
\label{sec:themodel}

The ionization code {\code} calculates the equilibrium electron density
and the ionization fractions of all ions, from which all ionic number
densities are computed in each grid cell.  Metals up to and including
zinc are incorporated, however, the user can select which metals are
included in the chemical mixture.  All ions are treated as two-level
systems, a bound ground state and the continuum; no recombination
transitions are incorporated.  Neither photo-heating nor cooling is
treated, since these are directly incorporated into hydroART and yield
the equilibrium temperature of the gas in the grid cells.
\citet{fumagalli11} examined the effects of additional heating by
artificially incrementing the temperatures of the grid cells and found
that the morphology of the neutral hydrogen gas was negligibly
modified for photoionized gas.

The equilibrium solution requires solving a matrix of coupled
non-linear rate equations.  We derived a method in which the rate
equations are linearized and the solution is obtained via iterative
convergence on the equilibrium ionization fractions using particle and
charge density conservation.

The physical gas processes included in {\code} are photoionization,
Auger ionization, direct collisional ionization, excitation
auto-ionization, charge exchange ionization, radiative recombination,
dielectronic recombination, and charge exchange recombination.  If
desired, the effects of each of these processes can be isolated by
turning the process ``off'' or ``on''.

\subsection{Notation and Formalism}
\label{sec:formalism}

For what follows, we denote the atomic species by the index $k$, where
$k$ equals the atomic number, and denote the ionization stage by the
index $j$, where $j=1$ is the neutral stage and $j=k+1$ is the fully
ionized stage.  We assume that all ions and neutral atoms are in their
ground state.  The number density [cm$^{-3}$] of ion $k,j$ is $n
\subkj$ and the electron number density is $n_{\hbox{\tiny e}}$.

The rate equation, $dn\subkj / dt$, quantifies the rate of change in
the number of ion $k,j$ per unit volume per unit time
[cm$^{-3}$~s$^{-1}$].  It can be expressed as 
\begin{equation}
\begin{array}{lcl}
\displaystyle \frac{dn\subkj}{dt} &=& 
(\hbox{creation rate ion $k,j$}) \,\,\,\, - \\[3pt]
& &  (\hbox{destruction rate of ion $k,j$}) \, .
\end{array}
\end{equation}
Clearly, there is a rate equation for each ionization stage for each
atomic species, which taken together form a rate matrix.  The rate
matrix is closed by particle and charge density conservation.

The creation and destruction rates per unit volume are determined by
multiplying the rate per unit time [s$^{-1}$] by the number density of
the initial state particle.  For example, in the case of
photoionization (denoted ``ph'') of ion $k.j$, the contribution to the
destruction rate per unit volume is $n\subkj R^{\hbox{\tiny
    ph}}\subkj$.  In the case of recombination (denoted ``rec'') with
ion $k,j-1$ to create ion $k,j$, the contribution to the creation rate
per unit volume is $n_{\hbox{\tiny k,j-1}}R^{\hbox{\tiny
    rec}}_{\hbox{\tiny k,j-1}}$.  Note that all rates, $R\subkj$, are
indexed to the initial ion stage.

All collision based rates are determined from the rate coefficients
[cm$^3$~s$^{-1}$].  We denote ionization rate coefficients as $\alpha
\subkj$ and recombination rate coefficients as $\beta \subkj$.  In the
case that ionization is due to a collision with a free electron, the
rate per unit time is obtained by multiplying the rate coefficient by
the electron density.  We use the convention that ionization rate
coefficients are indexed by referencing the initial ion stage $j$,
whereas, for recombination, $j$ refers to the final ion stage.  That
is, for recombination, rate coefficients index the ion towards which
recombination proceeds.  

For example, in the case of direct collisional ionization (denoted
``cdi'') of ion $k,j$ by a free electron, the contribution to the
destruction rate per unit volume of ion $k,j$ is $n\subkj
R^{\hbox{\tiny cdi}}\subkj = n\subkj \eden \alpha ^{\hbox{\tiny
cdi}}\subkj$.  In the case of radiative recombination (denoted
``phr'') of a free electron with ion $k,j$ to create ion $k,j-1$, the
contribution to the destruction rate per unit volume of ion $k,j$ is
$n\subkj R^{\hbox{\tiny phr}}\subkj = n\subkj \eden \beta
^{\hbox{\tiny phr}}\subkjm$.  Note that this expression would also be
the contribution to the radiative recombination creation rate per unit
volume for ion $k,j-1$

In Section~\ref{sec:rateeqs}, we write out the rate equations for all
ions.  The equilibrium balance is achieved when $dn \subkj /dt = 0$
for all ions, by balancing the creation and destruction rates per unit
volume. In Section~\ref{sec:solution}, we outline our method to solve
the rate matrix.

\subsection{Particle and Charge Density Conservation}
\label{sec:abundances}

Together, particle and charge density conservation provide the
constraints for obtaining the ion densities and ionization fractions.

The total number density of all ions and free electrons is
$n_{\hbox{\tiny tot}} = n_{\hbox{\tiny A}} + n \sube $, where $n\sube$
is the density of free electrons and, as dictated by particle
conservation,
\begin{equation}
n_{\hbox{\tiny A}}  
= {\textstyle \sum \limits _{\hbox{\tiny k=1}} n \subk } 
= {\textstyle \sum \limits _{\hbox{\tiny k=1}}
              \sum \limits _{\hbox{\tiny j=1}}^{\hbox{\tiny k+1}} n \subkj } \, ,
\label{eq:particleconservation}
\end{equation}
is the number density of all atomic species, where $n \subk$ is the
number densities of species $k$, and $n \subkj$ is the number density
of species $k$ in ionization stage $j$. Since the ionization model
includes elements only up to zinc, the maximum $k$ is 30.  And, since
the user can specify a subset of these elements for inclusion into the
ionization model, the sum includes only the $k$ values of the atomic
species used in the ionization model.


The abundance fractions, $\eta \subk = n\subk/n_{\hbox{\tiny A}}$, are
employed to compute the number density of the atomic species.  With
this formalism, the number density of each atomic species is
determined directly from the input hydrogen number density and
abundance fractions via $n\subk = \eta\subk n_{\hbox{\tiny A}} =
(\eta\subk/\eta\subH) n\subH$, where $\eta\subH \equiv
\eta_{\hbox{\tiny 1}}$ and $n\subH \equiv n_{\hbox{\tiny 1}}$.  The
$\eta \subk$ are determined by the mass fractions, $x\subk$, according
to
\begin{equation}
\eta \subk = \frac{x\subk/A\subk}{\textstyle \sum \limits
  _{\hbox{\tiny k=1}} x\subk/A\subk} \, ,
\label{eq:afracs}
\end{equation}
where $A\subk$ is the atomic mass of species $k$ in units of the
unified atomic mass unit, $m_{\hbox{\tiny a}} = 1.6605\times 10^{-24}$
[g].  Again, the sum includes only the elements used in the ionization
model in order to preserve $\sum _{\hbox{\tiny k=1}} \eta\subk = 1$.

In hydroART, each grid cell records the hydrogen density and the type
II and type Ia total mass fractions, $Z_{\hbox{\tiny II}}$ and
$Z_{\hbox{\tiny Ia}}$.  To compute the abundance fractions from
Eq.~\ref{eq:afracs}, we require the individual mass fractions of
hydrogen, helium, and the included metal species. For type II
composition, we use the production factors from the rotating models of
\citet{chieffi13}.  The production factors are determined by
integrating the model yields for different mass progenitors across the
Salpeter initial mass function \citep[$\alpha = 2.35$,][]{salpeter55}
over the progenitor mass range 13--120~M$_{\odot}$.  For type Ia
composition, we use the yields, $(Y\subk/Y_{\hbox{\tiny k$\odot$}})/
(Y_{\hbox{\tiny Fe}}/Y_{\hbox{\tiny Fe$\odot$}})$ from the C series
DD2 models of \citet{iwamoto99}, which are the most consistent with
observational constraints of the nucleosynthesis products in the
Galaxy.

We denote the type II and type Ia mass fractions for atomic species
$k$ as $(x\subk)_{\hbox{\tiny II}}$ and $(x\subk)_{\hbox{\tiny Ia}}$,
respectively.
For hydrogen and helium, $Y_{\hbox{\tiny 1}} = Y_{\hbox{\tiny 2}} = 0
$, yielding $(x_{\hbox{\tiny 1}})_{\hbox{\tiny Ia}} = (x_{\hbox{\tiny
    2}})_{\hbox{\tiny Ia}} = 0$.  The metal mass fractions are
rescaled by a constant to recover the mass fractions in the grid cell,
\begin{equation}
Z_{\hbox{\tiny II}} = C_{\hbox{\tiny II}} 
\textstyle \sum \limits _{\hbox{\tiny k=3}} (x\subk)_{\hbox{\tiny II}} \, ,
\qquad
Z_{\hbox{\tiny Ia}} = C_{\hbox{\tiny Ia}} 
\textstyle \sum \limits _{\hbox{\tiny k=3}} (x\subk)_{\hbox{\tiny Ia}} \, .
\end{equation}
For consistency, only the atomic species that are incorporated in the
ionization model are included in the sums.  We now need to rescale the
hydrogen and helium mass fractions.  Since all the hydrogen and helium
originates from the type II ejecta\footnote{The grid cells are
  initially given the primordial hydrogen and helium mass fractions at
  $z=50$, the starting redshift of the simulation.  We do not apply a
  chemical evolution model to account for the full history of the grid
  cells, but assume that the feedback enrichment dominates.}, we
define $r = (x_{\hbox{\tiny 2}})_{\hbox{\tiny II}}/(x_{\hbox{\tiny
    1}})_{\hbox{\tiny II}}$, and obtain the mass fractions for the
mixture (employed in Eq.~\ref{eq:afracs}),
\begin{equation}
x_{\hbox{\tiny 1}} = 
\frac{1-\left(\, Z_{\hbox{\tiny II}}+Z_{\hbox{\tiny Ia}} \, \right) }{1+r} 
\, , \quad
x_{\hbox{\tiny 2}} = rx_{\hbox{\tiny 1}}
\, , \quad
x\subk = 
 C_{\hbox{\tiny II}} (x\subk)_{\hbox{\tiny II}} + 
 C_{\hbox{\tiny Ia}} (x\subk)_{\hbox{\tiny Ia}}  \, ,  
\end{equation}
which preserves the constraint $\sum _{\hbox{\tiny
    k=1}} x\subk = 1$.  

If desired, solar abundance pattern can be used, in which case we
directly employ the solar mass fractions from Table 1.4 of
\citet{draine11}, which are derived from \citet{asplund09}.  Finally,
the metallicity of the gas in solar units is
\begin{equation}
Z/Z_{\odot} = \frac
{\textstyle \sum \limits _{\hbox{\tiny k=3}} (x\subk/x\subH)}
{\textstyle \sum \limits _{\hbox{\tiny k=3}} (x\subk/x\subH)_{\odot}}
\, ,
\end{equation}
where $x\subH = x_{\hbox{\tiny 1}}$.

Since each ion $k,j$ donates $j-1$ electrons to the free electron
pool, the contribution to the electron density from each ion is
$(j-1)n\subkj$.  The total electron density from all ions is then 
\begin{equation}
n \sube 
= {\textstyle \sum \limits _{\hbox{\tiny k}} 
    \sum \limits _{\hbox{\tiny j=2}}^{\hbox{\tiny k+1}} } (j-1) \, n\subkj \, .
\end{equation}
Introducing the ionization fractions, $f\subkj (n\sube ,T, J_{_E}) = n
\subkj / n\subk $, we write $n\subkj = f\subkj n\subk =
f\subkj \eta\subk n\subA$, and obtain the equation for charge density
conservation,
\begin{equation}
n \sube 
= n_{\hbox{\tiny A}} {\textstyle \sum \limits _{\hbox{\tiny k}} } \eta \subk 
  {\textstyle \sum \limits _{\hbox{\tiny j=2}}^{\hbox{\tiny k+1}}  } (j-1)  
   f\subkj (n\sube ,T, J_{_E}) \, .
\label{eq:chargeconservation}
\end{equation}
Note that the ionization fraction depends upon the electron density,
temperature, and the ionizing photon field,
$J_{_E}$. Eq.~\ref{eq:chargeconservation} is a linear transcendental
equation; the electron density, $n \sube$, must be known in order to
calculated the ionization fractions, which are constrained via
particle and charge density conservation to yield a free electron pool
with density $n \sube$.  

\subsection{Ionizing Spectrum}
\label{sec:seds}


We provide the options (1) ionization by the ultraviolet background
(UVB), which is redshift dependent, (2) ionization by stellar
populations, which depends on each populations total stellar mass,
age, metallicity, and distance from the model cloud, and (3)
ionization by both the UVB and stellar populations.

For the UVB, we use the SEDs of \citet{haardt11} added to the cosmic
microwave background.  These spectra are specified as the specific
intensity, $J_\nu$ [erg s$^{-1}$ cm$^{-2}$ str$^{-1}$ Hz$^{-1}$] over
the energy [eV] interval $-6.7 \leq \log E \leq 6.8$ and are provided
for redshifts $0 \leq z \leq 5$.  Once the cloud redshift is
specified, a grid of SEDS sampled at intervals of $\Delta z = 0.2$ are
cubic spline interpolated at each frequency to obtain $J_\nu (z)$,
which is then converted to the specific intensity per unit energy,
$J_{_E} (z)$ [cm$^{-1}$ s$^{-1}$ str$^{-1}$] versus $E$ [eV] via
$J_{_E} dE = J_\nu \, d\nu $.

For the stellar populations, we use SEDs computed from the Starburst99
v6.02 models \citep{Sb99-99}.  We built a library of SEDs comprising
stellar populations of $M = 10^3$, $10^4$, $10^5$, and
$10^6$~M$_\odot$.  For each mass, five ages were computed (1, 5, 10,
20, and 40 Myr) and for each mass and age five metallicities were
computed ($10^{-4}$, $10^{-3}$, $10^{-2}$, $10^{-1}$, and
$1~Z_\odot$).  We store the SEDs as the luminosity density per unit
wavelength, $L_{_\lambda}$ [erg s$^{-1}$~{\AA}$^{-1}$] over the wavelength
range $91 \leq \lambda \leq 1.6 \times 10^6$~{\AA}, which corresponds
to the energy [eV] interval $-2.1 \leq \log E \leq 2.1$.

Once the desired mass, age, and metallicity are specified, we first
cubic spline interpolate $L_{_\lambda}$ at each wavelength across mass
for each age and metallicity, then across age for each metallicity,
and then across metallicity.  The final SED is then converted to the
flux density per unit wavelength, $F_{_\lambda} = L_{_\lambda}/(4\pi
r^{\hbox{\tiny 2}})$ [erg s$^{-1}$ cm$^{-2}$ str$^{-1}$ {\AA}$^{-1}$],
where $r$ is the specified distance to the stellar population from the
model cloud. Finally, we convert the SED to $J_{_E}$ versus $E$.  If a
combined UVB plus stellar population SED is used, the two specific
intensities are added.  In general, the contribution by stars scales
linearly with the mass of the stellar population and with the inverse
square of the distance between the model cloud and the stellar
population.

\begin{figure}
\epsscale{1.2}
\plotone{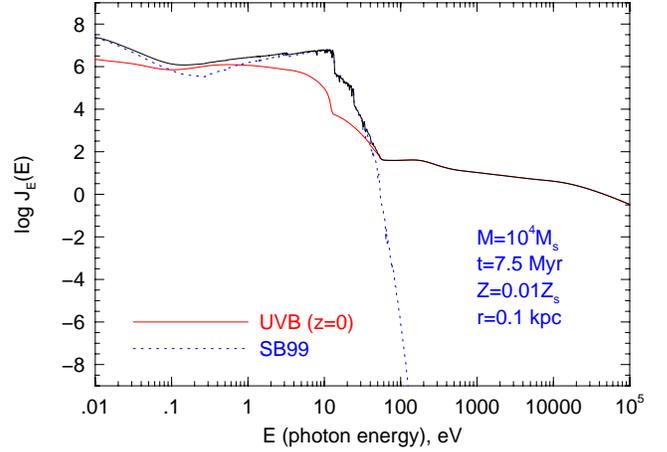}
\caption{An example spectral energy distribution, $J_{_E}$ [cm$^{-1}$
    s$^{-1}$ str$^{-1}$] versus $E$ [eV].  The red curve is the
  \citet{haardt11} UVB for $z=0$.  The blue dotted curve is the
  Starburst99 \citep{Sb99-99} model (SB99) for a single stellar
  population of mass $10^4$~M$_{\odot}$ with age 7.5~Myr, and a
  metallicity of 0.01 in solar units.  The black curve is the total of
  the two contributions.  For this example, the the stellar population
  is assumed to be at a distance of 100 pc from the model cloud.}
\label{fig:seds}
\end{figure}

In Figure~\ref{fig:seds}, we illustrate a SED that combines
contributions from both the UVB (red curve) and a stellar population
(blue dotted curve).  For this example, the UVB is a $z=0$
\citet{haardt11} spectrum and the stellar population has mass
$M_{\ast}=10^4$~M$_{\odot}$, age $t=7.5$ Myr, and metallicity $\log
Z/Z_{\odot} = -1$ and is at a distance of 100 pc.  There is no
attenuation.  Typical of this example, stellar radiation modifies the
UVB SED in the spectral region below 100 eV, and when present, will
generally lead to higher ionization conditions in regions where
photoionization dominates. Even with no assumed attenuation through
the ISM, stellar SEDs rarely modify the UVB incident on a grid cell
unless the stellar population is very close in proximity, otherwise
the population needs to be very young (populated with O stars) and
massive \citep{cl98}.

\subsection{Ionization Rates}
\label{sec:ionization}

We treat photoionizationa and Auger ionization
(Section~\ref{subsec:photoionization}), direct collisional ionization
(Section~\ref{sec:cdi}), and excitation auto-ionization processes
(Section~\ref{sec:cea}).  We also treat charge exchange ionization, which
is presented in Section~\ref{sec:chargeexchange}.

\begin{figure*}
\epsscale{1.15}
\plotone{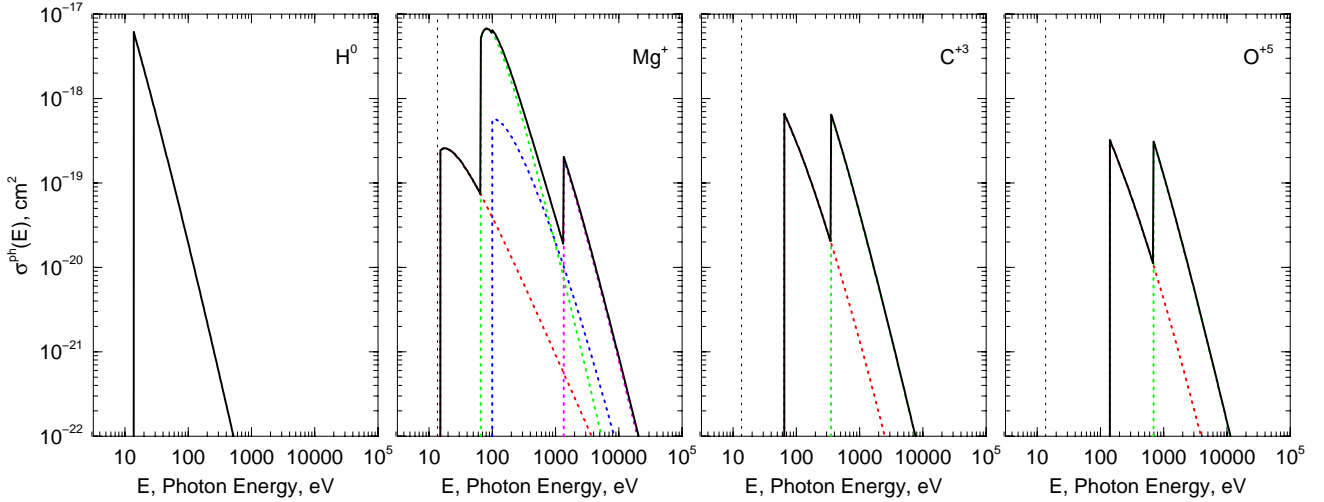}
\caption{The photoionization cross sections [cm$^2$] for
  H$^{\hbox{\tiny 0}}$, Mg$^{+}$, C$^{\hbox{\tiny +3}}$, and
  O$^{\hbox{\tiny +5}}$ as a function of the photon energy [eV].  For
  H$^{\hbox{\tiny 0}}$, the cross section is for the 1s shell.  For
  Mg$^{+}$, the red curve is the 3s shell, the green curve is the 2p
  shell, the blue curve is the 2s shell, and the magenta curve is the
  1s shell.  For C$^{\hbox{\tiny +3}}$ and O$^{\hbox{\tiny +5}}$, the
  red curve is the 2s shell and the green curve is the 1s shell.  The
  total cross sections are given by the black curves.  For reference,
  the vertical dotted line is the ground-state ionization energy for
  H$^{\hbox{\tiny 0}}$.}
\label{fig:phxsecs}
\end{figure*}

\subsubsection{Photo and Auger Ionization}
\label{subsec:photoionization}

Photoionization and Auger ionization both begin with radiative
ionization processes in which a bound electron is photo-ejected from
an ion or neutral atom.  The electron can be liberated from any one of
the ion's populated electron shells.

To compute the rate, $R^{\, \hbox{\tiny ph}} \subkjs$, at which an
electron in a given shell, index $s$, is photo-ejected from ion $k,j$,
we multiply the cross section for absorption, $\sigma \ph \subkjs
(E)$, for shell $s$ at energy $E=h\nu$ by the photon number density
per unit energy, $4\pi J_{_E}/E$ [erg$^{-1}$ cm$^{-2}$~s$^{-1}$], and
integrate over all energies greater than the binding energy of the
electron,
\begin{equation}
R^{\, \hbox{\tiny ph}} \subkjs = 
4 \pi \int ^{\infty}_{I \subkjs} 
J_{_E} \sigma \ph \subkjs (E) \, \frac{dE}{E} \, ,
\label{eq:Rphoto}
\end{equation}
where $I \subkjs$ is the ionization (binding) energy for electrons in
shell $s$ of ion $k,j$.

For the computation of Eq.~\ref{eq:Rphoto}, we adopt the convention
for the shell indices such that $s=1$ is the 1s shell, $s=2$ is 2s,
$s=3$ is 2p, $s=4$ is 3s, $s=5$ is 3p, $s=6$ is 3d, and $s=7$ is 4s.
The photoionization cross sections are computed from the fitting
functions and fitting parameters tabulated by \citet{verner95} for
inner shells and by \citet{verner96} for the outer shells.  Their work
includes all ionization stages and shells for hydrogen through zinc.
We computed $\sigma \ph \subkjs (E)$ from,
\begin{equation}
\sigma \ph \subkjs (E) = \sigma_{_0} \, y^{_{\, -Q}} \, \frac{(x-1)^2
  + y^2_{\hbox{\tiny W}}}{[1 + (y/y_{\hbox{\tiny A}})^{\hbox{\tiny
        1/2}}]^{^P}} \, ,
\label{eq:phxsec}
\end{equation}
where $x = E/E_{_0} - y_{_0}$ and $y = (x^{\hbox{\tiny 2}} +
y^{\hbox{\tiny 2}}_{_1})^{\hbox{\tiny 1/2}}$.  The tabulated fitting
parameters for each $k,j,s$ are $\sigma_{_0}$, $E_{_0}$,
$y_{\hbox{\tiny A}}$, $P$, $y_{\hbox{\tiny W}}$, $y_{_0}$, and
$y_{_1}$.  For inner shells $y_{\hbox{\tiny W}}$, $y_{_0}$, and
$y_{_1}$ are null and the asymptotic power is $Q =
\frac{1}{2}P+\ell+\frac{11}{2}$, where $\ell$ is the angular momentum
quantum number of the shell.  For the outer shell $Q =
\frac{1}{2}P+\frac{11}{2}$.  The physical interpretation of each
fitting parameter is explained in \citet{verner96}.

In Figure~\ref{fig:phxsecs}, we present the photoionization cross
sections for H$^{\hbox{\tiny 0}}$, and Mg$^{+}$, and C$^{\hbox{\tiny
+3}}$, and O$^{\hbox{\tiny +5}}$ as a function of the photon energy as
computed from Eq.~\ref{eq:phxsec} and applied in Eq.~\ref{eq:Rphoto}.
Ground-state Mg$^{+}$ has the isoelectronic sequence of neutral sodium
(1s$^2$2s$^2$2p$^6$3s$^1$) and ground-state C$^{\hbox{\tiny +3}}$, and
O$^{\hbox{\tiny +5}}$ have the isoelectronic sequence of neutral
lithium (1s$^2$2s$^1$).  The individual shell cross sections are shown
as colored dotted curves and the total is the solid curve.  The red
curves are the ground-state ionization threshold energies for the
outermost populated electron shell.

In the case of photoionization, a single electron, ${\rm
  e}_{\hbox{\tiny ej}}^-$, is ejected and the ionization stage of the
incident ion $k,j$ is incremented by one to $k,j+1$,
\begin{equation}
A \subkj + \gamma \rightarrow 
A _{\hbox{\tiny k,j+1}} + {\rm e}_{\hbox{\tiny ej}}^- \, .
\end{equation}
In cases where the incident photon has the required energy to liberate
an inner shell electron some of its energy can be channeled into
also liberating one or more of the less bound, higher shell electrons.
This process is known as Auger ionization, in which the ionization
stage of the incident ion $k,j$ is incremented by two or more,
\begin{equation}
A \subkj + \gamma \rightarrow 
A _{\hbox{\tiny k,m}} + (m-j) \, {\rm e}_{\hbox{\tiny ej}}^- \, ,
\end{equation}
where we use the convention that that the initial ionization stage is
$j$ and the final higher ionization stage is $m$.  Note that the
number of ejected electrons is $N_{\rm e} = m - j$.  Because the
photo-electron is included in the notation, the final stage $m$ is
always greater than or equal to $j+2$.  Photoionization is the special
case in which $m = j +1$.

In order to compute the photo and Auger ionization rates, it is
necessary to know the yield probability, $P^{\,\, \hbox{\tiny
s}}_{\hbox{\tiny k,j,m-j}}$, i.e., the probability that $N_{\rm e} = m
- j$ electrons in total are ejected from an ion following a
photoionization of an electron originating from shell $s$ (the
photo-electron).  These yield probabilities have been calculated and
tabulated by \citet{kaastra93} for each electron shell.  

For photoionization, the total photoionization rate, $R^{\,
  \hbox{\tiny ph}}_{\hbox{\tiny k,j}}$, for destruction of ion $k,j$
is given by $R^{\, \hbox{\tiny ph}} \subkjs$ (see
Eq.~\ref{eq:Rphoto}), the rate at which an electron bound in shell $s$
of ion $k,j$ is ionized by incident photons, weighted by the
probability that only the photo-electron is ejected from the ion and
summed over all electron shells,
\begin{equation}
R^{\, \hbox{\tiny ph}}_{\hbox{\tiny k,j}} = 
\textstyle \sum \limits_{s=1} ^{N^{\, \hbox{\tiny s}}_{\hbox{\tiny k,j}}}
P^{\,\, \hbox{\tiny s}}_{\hbox{\tiny k,j,1}} R^{\,\, \hbox{\tiny ph}} \subkjs  \, ,
\label{eq:Rphcalc}
\end{equation}
where $N^{\, \hbox{\tiny s}}_{\hbox{\tiny k,j}}$ is the number of shells
for ion $k,j$.  

\begin{figure*}
\epsscale{1.15}
\plotone{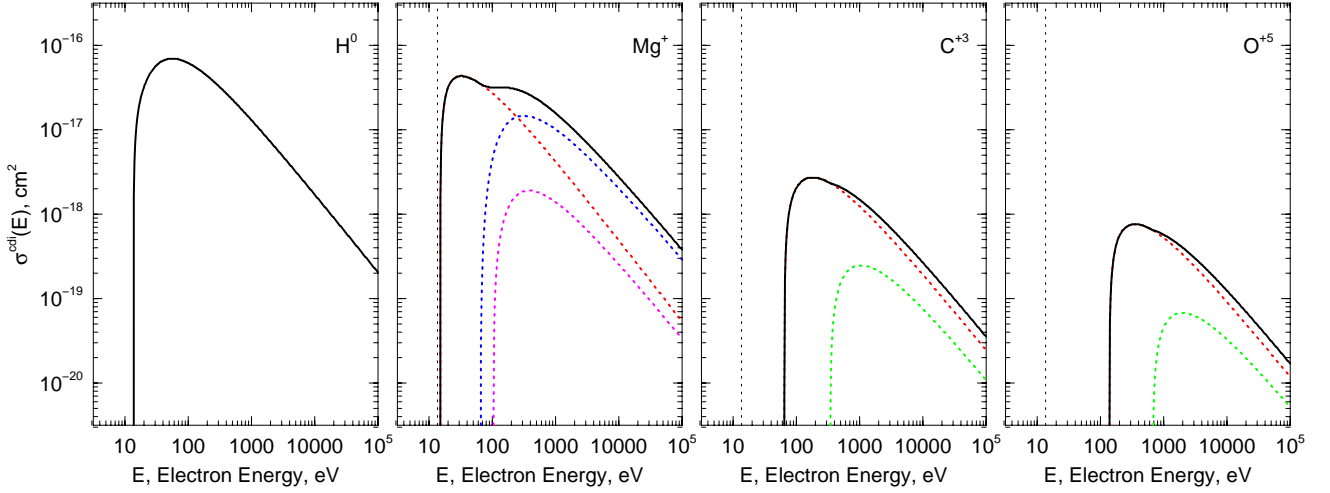}
\caption{The direct collisional ionization cross sections [cm$^2$] for
  H$^{\hbox{\tiny 0}}$, Mg$^{+}$, C$^{\hbox{\tiny +3}}$, and
  O$^{\hbox{\tiny +5}}$ as a function of the electron energy [eV].
  For H$^{\hbox{\tiny 0}}$, the cross section is for the 1s shell.
  For Mg$^{+}$, the red curve is the 3s shell, the blue curve is the
  2s+2p shell, and the magenta curve is the 1s shell.  For
  C$^{\hbox{\tiny +3}}$ and O$^{\hbox{\tiny +5}}$, the red curve is
  the 2s shell and the green curve is the 1s shell.  The total cross
  sections are given by the black curves.  For reference, the
  vertical dotted line is the ground-state ionization energy for
  H$^{\hbox{\tiny 0}}$.}
\label{fig:cdixsecs}
\end{figure*}

Similarly, the Auger ionization rate, $R^{\, \hbox{\tiny
aug}}_{\hbox{\tiny k,j,m}}$, for destruction of an ion $k,j$ that
ejects $N_{\rm e} = m - j$ electrons (including the photo-electron, so
$N_{\rm e} \geq 2$) is the sum of $R^{\, \hbox{\tiny ph}} \subkjs$ over
all electron shells weighted by the probability that $N_{\rm e}$
electrons in total were ejected from the ion in response to a
photo-electron originating in shell $s$,
\begin{equation}
R^{\, \hbox{\tiny aug}}_{\hbox{\tiny k,j,m}} = 
\textstyle \sum \limits_{s=1} ^{N^{\hbox{\tiny s}}_{\hbox{\tiny k,j}}}
P^{\,\, \hbox{\tiny s}}_{\hbox{\tiny k,j,m-j}} R^{\, \hbox{\tiny ph}} \subkjs  \, .
\end{equation}
The $R^{\, \hbox{\tiny aug}}_{\hbox{\tiny k,j,m}}$ are unique amongst
the various ionization rates, because they dictate the balance between
non-adjacent ionization stages of species $k$.  Clearly, $R^{\,
  \hbox{\tiny aug}}_{\hbox{\tiny k,k+1,m}} = R^{\, \hbox{\tiny
    aug}}_{\hbox{\tiny k,k,m}} = 0$ since fully ionized and hydrogenic
ions cannot undergo Auger ionization.

\subsubsection{Direct Collisional Ionization}
\label{sec:cdi}

Direct collisional ionization is the collision of an electron with an
ion, which then directly ionizes from $j$ to $j+1$,  
\begin{equation}
A \subkj + {\rm e}_{\hbox{\tiny f}}^- \rightarrow A _{\hbox{\tiny
k,j+1}} + {\rm e}_{\hbox{\tiny f}}^- + {\rm e}_{\hbox{\tiny ej}}^- \, .
\end{equation}
In the process, the colliding free electron, ${\rm e}_{\hbox{\tiny
f}}^-$, loses an energy equal to the ionization energy plus the
kinetic energy of the ejected electron.

The ionization rate for destruction of ion $k,j$ due to direct
collisional ionization is obtained by multiplying the total direct
collisional ionization rate coefficient, $\alpha^{\hbox{\tiny
cdi}}_{\hbox{\tiny k,j}} (T)$, by the number density of free electrons,
\begin{equation}
R \subkj \cdi (T) = \eden \alpha^{\hbox{\tiny cdi}}_{\hbox{\tiny k,j}} (T) \, ,
\end{equation}
where 
\begin{equation}
\alpha \cdi \subkj (T) = 
\textstyle \sum \limits_{s=1} ^{N^{\hbox{\tiny s}}_{\hbox{\tiny k,j}}}
\alpha \cdi \subkjs (T) \, .
\label{eq:alpchcditot}
\end{equation}
is the sum of the direct collisional ionization rate coefficient
contributions, $\alpha \cdi \subkjs (T)$, for ejection of an electron
from shell $s$.  Here, the shell indices are $s=1$ is the 1s shell,
$s=2$ is the combined 2s+2p shell, $s=3$ is the combined 3s+3p shell,
and $s=4$ is the 4s shell.

The $\alpha^{\hbox{\tiny cdi}}_{\hbox{\tiny k,j,s}} (T)$ are the
expectation values 
$ \left< \sigma \cdi \subkjs \cdot v \right> $, 
where $\sigma \cdi \subkjs (E)$ is the direct collisional
ionization cross section for the shell, $v(E) = \sqrt{2kE/m\sube}$ is
the electron speed for kinetic energy $E$, and $m\sube$ is the
electron mass. The integration is over energies large enough to
overcome the binding energy,
\begin{equation}
\begin{array}{lcl}
\alpha \cdi \subkjs (T) \!\!\! & = & \!\!\!
\left< \sigma \cdi \subkjs \cdot v \right> 
\\[10pt]
 \!\!\! & = & \!\!\! \displaystyle \sqrt{\frac{2k}{m\sube}} 
\int _{I\subkjs}^{\infty} \!\!\!\! 
\sigma \cdi \subkjs (E) f(E,T)\, E^{\hbox{\tiny 1/2}} dE \, ,
\end{array}
\label{eq:alphacdicomp}
\end{equation}
where $f(E,T)$ is the Maxwell-Boltzmann speed distribution function at
equilibrium temperature $T$, and $I\subkjs$ is the ionization energy
of shell $s$ of ion $k,j$, .

The direct collisional ionization cross sections and rate coefficients
are computed from the fitting functions and parameters tabulated by
\citet{arnaud85}.  For shell $s$ of ion $k,j$, let $u_{_1} =
E/I\subkjs$ and $u_{_2} = 1 - I\subkjs /E$. The direct collisional
ionization cross section for shell $s$ is computed from
\begin{equation}
\sigma \cdi \subkjs (E) = 
\frac{10^{-14}}{u_{_1} I \subkjs ^{\hbox{\tiny 2}}} 
\left\{ 
a \jks u_{_2} + 
b \jks u_{_2}^{\hbox{\tiny 2}} +
c \jks \ln u_{_1} +
d \jks \frac{\ln u_{_1}}{u_{_1}} 
\right\} \, ,
\label{eq:xsec-cdi}
\end{equation}
where the four fitting coefficients, $a \jks$, $b \jks$, $c \jks$, and
$d \jks$, are tabulated in \citet{arnaud85} for each shell for all ion
stages of hydrogen through nickel.  The units of the fitting
coefficients are $10^{-14}$ cm$^2$ eV$^2$.  In
Figure~\ref{fig:cdixsecs}, we present the direct collisional
ionization cross sections for H$^{\hbox{\tiny 0}}$, and Mg$^{+}$, and
C$^{\hbox{\tiny +3}}$, and O$^{\hbox{\tiny +5}}$ as a function of the
electron energy as computed from Eq.~\ref{eq:xsec-cdi}.  The
individual shell cross sections are shown as colored dotted curves.
The red curves are the ground-state ionization threshold energies for
the outermost populated electron shell.

Rather than integrate Eq.~\ref{eq:alphacdicomp} using
Eq.~\ref{eq:xsec-cdi}, which can be computationally expensive, we
computed $\alpha \cdi \subkjs (T)$ for each shell using the fitting
formulae of \citet{arnaud85}, for which the same fitting coefficients
employed for the cross sections apply.  Let $x \jks = I \subkjs / kT$,
then
\begin{equation}
\alpha \cdi \subkjs (T) = 
\frac{6.69 \times 10^{-7}}{\left( kT \right) ^{\hbox{\tiny 3/2}}}
F\jks (x \jks) \frac{\exp \left\{ - x \jks \right\}}{x \jks} \, ,
\end{equation}
where
\begin{equation}
\begin{array}{lcl}
\displaystyle
F \jks (x) \!\! & = & \!\!
a \jks \left[1 - x f_{_1} (x) \right]  +  
b \jks \left[1 + x  - x (2+  x ) f_{_1}(x) \right] \\[10pt]
& & \!\! + \, \, 
c \jks f_{_1}(x ) + d \jks \, x  f_{_2}(x) \,
\end{array}
\end{equation}
and where
\begin{equation}
f_{_1}(x) = e^{x} \int _1^{\infty} \frac{dt}{t} e^{-xt}
\qquad 
f_{_2}(x) = e^{x} \int _1^{\infty} 
\frac{dt}{t} e^{-xt} \ln t \, .
\label{eq:f1x}
\end{equation}
The integral for $f_{_1}(x)$ is the well known Exponential function.
Both $f_{_1}(x)$ and $f_{_2}(x)$ are computed from the closed form
formulae given in \citet{arnaud85}, incorporating the corrections
given by \citet{verner90}.  We then compute the total direct
collisional ionization rate coefficient, $\alpha \cdi \subkj (T)$, for
ion $k,j$ using Eq.~\ref{eq:alpchcditot}.  We also examined the rate
coefficient formulae and parameters of \citet{voronov97}, but found
the results comparable.

\subsubsection{Excitation-Auto Ionization}
\label{sec:cea}

Excitation auto-ionization (E-A) occurs in ions with many inner filled
shell electrons and only a few outer shell electrons.  A collision
with a free electron first excites the ion. Then, during the internal
de-excitation process the released energy can either channel into
recombination emission lines or into liberating an outer shell
electron, which is auto-ionization,
\begin{equation}
A \subkj + {\rm e}_{\hbox{\tiny f}}^- 
\rightarrow A ^{\ast}_{\hbox{\tiny k,j}} + {\rm e}_{\hbox{\tiny f}}^- 
\rightarrow A _{\hbox{\tiny k,j+1}}  +  {\rm e}_{\hbox{\tiny ej}}^- 
+ {\rm e}_{\hbox{\tiny f}}^-\, ,
\end{equation}
where ${\rm e}_{\hbox{\tiny f}}^-$ the free collisional electron and
the ${\rm e}_{\hbox{\tiny ej}}^-$ is ejected auto-ionized electron.

The ionization rate for destruction of ion $k,j$ due to E-A
collisional ionization is obtained by multiplying the E-A
collisional ionization rate coefficient, $\alpha^{\hbox{\tiny
cea}}_{\hbox{\tiny k,j}} (T)$, by the number density of free electrons,
\begin{equation}
R \subkj \cea (T) = \eden \alpha^{\hbox{\tiny cea}}_{\hbox{\tiny k,j}} (T) \, .
\end{equation}

The total E-A rate coefficient is given by the expectation value $
\alpha \cea \subkj (T) = \left< \sigma \cea \subkj \cdot v \right> $
computed from Eq.~\ref{eq:alphacdicomp} with $I\subkjs$ replaced by
$\chi \subkj$, the E-A onset energy and $\sigma \cdi \subkjs (E)$
replaced by $\sigma \cea \subkj (E)$, the total E-A cross section,
which is obtained by weighting the de-excitation transition channels
over all transitions.

The E-A cross sections and rate coefficients depend on the bound
electron configuration of the ion, i.e., the isoelectronic sequence.
For example, C$^{\hbox{\tiny +3}}$, N$^{\hbox{\tiny +4}}$, and
O$^{\hbox{\tiny +5}}$ all have the electron configuration of neutral
lithium (1s$^{\hbox{\tiny 2}}$2s) and are thus lithium isosequence
ions ($N_{\rm e} = 3$).  

There is no E-A process for hydrogen and helium sequence ions.  For
lithium sequence ions, the dominant contribution to the cross section
is the 1s--2p transition.  As the charge of the ion, $Z$, increases,
the branching ratio to E-A decreases.  No significant E-A contribution
to the direct collisional cross section is observed for the beryllium
sequence (except perhaps O$^{\hbox{\tiny +4}}$, which is neglected),
nor for the sequences from boron to neon, which differ only in the
number of 2p shell electrons. For the sodium sequence ($[{\rm
Ne}]$3s$^{\hbox{\tiny 1}}$), up to 18 transitions can contribute to
E-A, for which the relative importance increases with $Z$.  The
sequences from magnesium to argon differ in the number of 3p shell
electrons, and the relative importance of E-A decreases as the shell
fills.

\begin{figure}
\epsscale{1.0}
\plotone{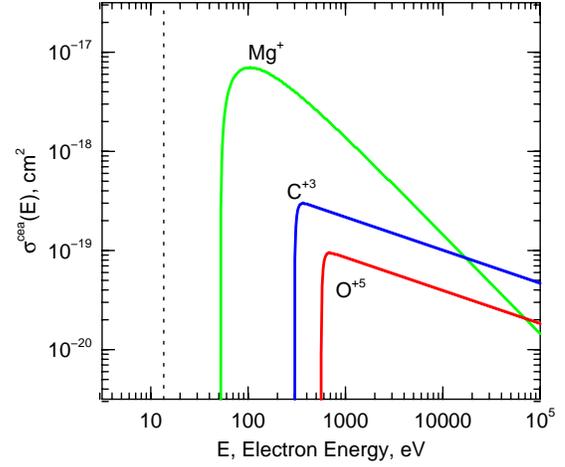}
\caption{The excitation auto-ionization collisional cross sections
  [cm$^2$] for Mg$^{+}$, C$^{\hbox{\tiny +3}}$, and
  O$^{\hbox{\tiny +5}}$ as a function of the electron energy [eV].
  For H$^{\hbox{\tiny 0}}$, the cross section is null.  Mg$^{+}$ is
  shown as the green curve, C$^{\hbox{\tiny +3}}$ as the blue curve,
  and O$^{\hbox{\tiny +5}}$ as the red curve. For reference, the
  vertical dotted line is the ground-state ionization energy for
  H$^{\hbox{\tiny 0}}$.}
\label{fig:ceaxsecs}
\end{figure}

\begin{deluxetable*}{lrclcccc}
\tablecolumns{8}
\tablewidth{0pt}
\tablecaption{Fitting Parameters for Excitation-Autoionization Cross Sections\tablenotemark{a} \label{tab:ceaxsecs}}
\tablehead{
\colhead{Sequence}      & 
\colhead{$N_{\rm e}$}   &     
\colhead{$Z$ Range}     &     
\colhead{$\chi \subkj$ [eV]}        &
\colhead{$Z_{\hbox{\tiny eff}}$} &
\colhead{$\sigma_{_0}$ [cm$^2$]} &
\colhead{$a$}           &
\colhead{$n$}           
}
\startdata
$[{\rm He}]$2s$^{\hbox{\tiny 1}}$ Li\tablenotemark{b} & 3  &  4--28 & $13.6\left\{(Z-0.835)^{\hbox{\tiny 2}} + (Z-1.62)^{\hbox{\tiny 2}} \right\} $ & $Z^{\hbox{\tiny 2}}$ & $4.22\times 10^{-16}$ & $1/3$ & 20 \\
$[{\rm Ne}]$3s$^{\hbox{\tiny 1}}$ Na (low $Z$)  & 11 & 12--16 & $26.0(Z-10)$                       & $(Z-11)^{\hbox{\tiny 0.35}}$ & $2.8\times 10^{-17}$ & $1$ & 1 \\
$[{\rm Ne}]$3s$^{\hbox{\tiny 1}}$ Na (high $Z$) & 11 & 18--28 & $11.0(Z-10)^{\hbox{\tiny 1.50}}$   & $(Z-10)^{\hbox{\tiny 1.87}}$ & $1.3\times 10^{-14}$ & $1$ & 3 \\
$[{\rm Ne}]$3s$^{\hbox{\tiny 2}}$ Mg & 12 & 18--28 & $10.3(Z-10)^{\hbox{\tiny 1.52}}$   & $Z$             & $4.0\times 10^{-13}/\chi \subkj$ & $1$ & 3 \\
$[{\rm Ne}]$3p$^{\hbox{\tiny 1}}$ Al & 13 & 18--28 & $18.0(Z-11)^{\hbox{\tiny 1.33}}$   & $Z$             & $4.0\times 10^{-13}/\chi \subkj$ & $1$ & 3 \\
$[{\rm Ne}]$3p$^{\hbox{\tiny 2}}$ Si & 14 & 18--28 & $18.4(Z-12)^{\hbox{\tiny 1.36}}$   & $Z$             & $4.0\times 10^{-13}/\chi \subkj$ & $1$ & 3 \\
$[{\rm Ne}]$3p$^{\hbox{\tiny 3}}$ P  & 15 & 18--28 & $23.7(Z-13)^{\hbox{\tiny 1.29}}$   & $Z$             & $4.0\times 10^{-13}/\chi \subkj$ & $1$ & 3 \\
$[{\rm Ne}]$3p$^{\hbox{\tiny 4}}$ S  & 16 & 18--28 & $40.0(Z-14)^{\hbox{\tiny 1.10}}$   & $Z$             & $4.0\times 10^{-13}/\chi \subkj$ & $1$ & 3 \\
\cutinhead{Special Cases}
$[{\rm Ar}]$4s$^{\hbox{\tiny 2}}$ Ca$^{\hbox{\tiny 0}}$                        & 20 & 20 & 25.0 & $\cdots $  & $6.0\times 10^{-17}$ & $1.12$ & $\cdots $ \\
$[{\rm Ar}]$4s$^{\hbox{\tiny 1}}$ Ca$^{+}$                       & 19 & 20 & 29.0 & $\cdots $  & $9.8\times 10^{-17}$ & $1.12$ & $\cdots $ \\
$[{\rm Ar}]$3d$^{\hbox{\tiny 3}}$4s$^{\hbox{\tiny 2}}$ Fe$^{\hbox{\tiny +3}}$  & 23 & 26 & 60.0 & $\cdots $  & $1.8\times 10^{-17}$ & $1.0$  & $\cdots $ \\
$[{\rm Ar}]$3d$^{\hbox{\tiny 2}}$4s$^{\hbox{\tiny 2}}$ Fe$^{\hbox{\tiny +4}}$  & 22 & 26 & 73.0 & $\cdots $  & $5.0\times 10^{-18}$ & $1.0$  & $\cdots $
\enddata
\tablenotetext{a}{The connection between isoelectronic series and index $k,j$ is $N_{\rm e} = k - j + 1$ and $Z=k$.}
\tablenotetext{b}{The fitting parameters for lithium are taken from \citet{hu96}.}

\end{deluxetable*}

\begin{deluxetable*}{lcccl}
\tablecolumns{5}
\tablewidth{0pt}
\tablecaption{Fitting Parameters for Excitation-Autoionization Rate Coefficients\tablenotemark{a} \label{tab:cearates}}
\tablehead{
\colhead{Sequence}      & 
\colhead{$Z_{\hbox{\tiny eff}}$} &
\colhead{$\alpha_{_0}$ [cm$^3$~s$^{-1}$~eV$^{1/2}$]} &
\colhead{$b$}           &
\colhead{$G_{\hbox{\tiny iso}}(x)$ : $a_{_0};a_{_1};a_{_2};a_{_3};a_{_4};a_{_5};a_{_6};a_{_7}$}
}
\startdata
$[{\rm He}]$2s$^{\hbox{\tiny 1}}$ Li\tablenotemark{b}  & $(Z-0.43)$ & $1.600\times 10^{-7}$ & $0.0002Z^{\hbox{\tiny 3}}$ & $0.67;1.20;0;0;2.22;-0.18;-1.20;0$ \\
$[{\rm Ne}]$3s$^{\hbox{\tiny 1}}$ Na  (low $Z$)        & $(Z-11)^{\hbox{\tiny 0.35}}$ & $1.873\times 10^{-9}/\chi \subkj$ & 0  & $1.00;0;0;0;0;-1.00;0,0$ \\
$[{\rm Ne}]$3s$^{\hbox{\tiny 1}}$ Na  (high $Z$)       & $(Z-10)^{\hbox{\tiny 1.87}}$ & $8.697\times 10^{-7}/\chi \subkj$ & 0 & $1.00;-0.50;0.50;0;0;0;0;-0.50$ \\
$[{\rm Ne}]$3s$^{\hbox{\tiny 2}}$ Mg                   & $Z$                          & $2.676\times 10^{-5}$             & 0                & $1.00;-0.50;0.50;0;0;0;0;-0.50$ \\
$[{\rm Ne}]$3p$^{\hbox{\tiny 1}}$ Al                   & $Z$                          & $2.676\times 10^{-5}$             & 0                & $1.00;-0.50;0.50;0;0;0;0;-0.50$ \\
$[{\rm Ne}]$3p$^{\hbox{\tiny 2}}$ Si                   & $Z$                          & $2.676\times 10^{-5}$             & 0                & $1.00;-0.50;0.50;0;0;0;0;-0.50$ \\
$[{\rm Ne}]$3p$^{\hbox{\tiny 3}}$ P                    & $Z$                          & $2.676\times 10^{-5}$             & 0                & $1.00;-0.50;0.50;0;0;0;0;-0.50$ \\
$[{\rm Ne}]$3p$^{\hbox{\tiny 4}}$ S                    & $Z$                          & $2.676\times 10^{-5}$             & 0                & $1.00;-0.50;0.50;0;0;0;0;-0.50$ \\
\cutinhead{Special Cases}
$[{\rm Ar}]$4s$^{\hbox{\tiny 2}}$ Ca$^{\hbox{\tiny 0}}$                          & 1.0  & $4.014\times 10^{-9}/\chi \subkj$ & 0 & $1.00;0;0;0;1.12;0;0;0$ \\
$[{\rm Ar}]$4s$^{\hbox{\tiny 1}}$ Ca$^{+}$                         & 1.0  & $6.556\times 10^{-9}/\chi \subkj$ & 0 & $1.00;0;0;0;1.12;0;0;0$ \\
$[{\rm Ar}]$3d$^{\hbox{\tiny 3}}$4s$^{\hbox{\tiny 2}}$ Fe$^{\hbox{\tiny +3}}$    & 1.0  & $1.204\times 10^{-9}/\chi \subkj$ & 0 & $1.00;0;0;0;1.00;0;0;0$ \\
$[{\rm Ar}]$3d$^{\hbox{\tiny 2}}$4s$^{\hbox{\tiny 2}}$ Fe$^{\hbox{\tiny +4}}$    & 1.0  & $3.345\times 10^{-9}/\chi \subkj$ & 0 & $1.00;0;0;0;-1.00;0;0;0$ 
\enddata
\tablenotetext{a}{The connection between isoelectronic series and index $k,j$ is $N_{\rm e} = k - j + 1$ and $Z=k$.}
\tablenotetext{b}{The tabulated value of $\alpha_{_0}$ for the lithium
sequence requires an additional multiplicative term.  
For C$^{\hbox{\tiny +3}}$, multiply by 0.6.
For N$^{\hbox{\tiny +4}}$, multiply by 0.8.
For O$^{\hbox{\tiny +5}}$, multiply by 1.25.
For all other ions, multiply by 1.2.}
\end{deluxetable*}

We computed the total E-A cross sections and rate coefficients for
ions up to nickel using the fitting functions and parameters tabulated
by \citet{arnaud85}, with the exceptions of the iron ions, which are
computed from the fitting functions and parameters updated by
\citet{arnaud92}, and the cross sections for the lithium sequence,
which we obtained from \citet{hu96}.




In order to simplify the individually presented fitting functions of
\citet{arnaud85} and \citet{hu96}, we present uniformly generalized
fitting functions for which we have distilled several of the their
fitting parameters into fewer terms.  For ion $k,j$, the total E-A
cross section is computed from
\begin{equation}
\sigma \cea \subkj (E) = 
\frac{\sigma_{_0}}{Z_{\hbox{\tiny eff}}^{\, ^2} } 
\frac{1}{u^a \subkj} \left( 1 - \frac{1}{u \subkj ^n} \right) \, ,
\label{eq:xsec-cea}
\end{equation}
where $u \subkj = E / \chi \subkj $, and where $E$ is the incident
electron kinetic energy.  The range of applicable $Z$, and the fitting
parameters, $ \chi \subkj$, $Z_{\hbox{\tiny eff}}$, $\sigma_{_0}$,
$a$, and $n$ are listed in upper panel of Table~\ref{tab:ceaxsecs} as
a function of isoelectronic sequence, given by $N_{\rm e}$. The
translation between isoelectronic sequence and the ion index $k,j$ is
given by $N_{\rm e} = k - j + 1$.  Four special cases are treated, for
which the fitting function takes the form
\begin{equation}
\sigma \cea \subkj (E) = 
\frac{\sigma_{_0}}{u\subkj} \left( 1 - a \ln u\subkj \right) \, ,
\end{equation}
and for which the fitting parameters are listed in the lower panel of
Table~\ref{tab:ceaxsecs}.  In Figure~\ref{fig:ceaxsecs}, we present
the E-A cross sections for Mg$^{+}$ (green curve), C$^{\hbox{\tiny
    +3}}$ (blue curve), and O$^{\hbox{\tiny +5}}$ (red curve) as a
function of the electron energy as computed from
Eq.~\ref{eq:xsec-cea}.

For ion $k,j$, the total E-A rate coefficients are computed from
\begin{equation}
\alpha \cea \subkj (T) = \frac{\alpha_{_0}}{Z_{\hbox{\tiny eff}}^{\,
^2} \, (1+b) } \, \frac{G_{\hbox{\tiny iso}}(x \subkj )}{\left( kT
\right) ^{\hbox{\tiny 1/2}}} \exp \{ - x\subkj \} \, ,
\label{eq:ceaalpha}
\end{equation}
where $x \subkj = \chi \subkj / kT$, and 
\begin{equation}
G_{\hbox{\tiny iso}}(x) = \textstyle \sum \limits _{n=0}^{3} 
     a_{\hbox{\tiny n}} x^{\hbox{\tiny n}} \, 
+ \, f_{_1}(x) \sum \limits _{n=0}^{3} a_{\hbox{\tiny n+4}} x^{\hbox{\tiny n}} \, ,
\end{equation}
where $f_{_1}(x)$ is given by Eq.~\ref{eq:f1x}.  The fitting
parameters, $Z_{\hbox{\tiny eff}}$, $\alpha_{_0}$, $b$, and
coefficients $a_{\hbox{\tiny n}}$ for $G_{\hbox{\tiny iso}}(x)$ are
listed in Table~\ref{tab:cearates}.  Note that the
range of applicable $Z$ and the ionization potentials used in
Eq.~\ref{eq:ceaalpha} are listed in Table~\ref{tab:ceaxsecs}.



\subsection{Recombination Rates}
\label{sec:recombination}

We treat radiative recombination (Section~\ref{sec:phr}) and
dielectronic recombination (Section~\ref{sec:die}).  Charge exchange
recombination is discussed in Section~\ref{sec:chargeexchange}.

\subsubsection{Radiative Recombination}
\label{sec:phr}

Radiative recombination is the capture of a free electron by ion
$k,j+1$ followed by the emission of a photon,
\begin{equation}
A \subkjp + {\rm e}_{\hbox{\tiny f}}^- \rightarrow A _{\hbox{\tiny
k,j}} + \gamma \, .
\end{equation}

The radiative recombination rate for creation of ion $k,j$ due to
electron recombination with ion $k,j+1$ is obtained by multiplying the
total recombination rate coefficient, $ \beta \phr \subkj (T)$, by the
electron number density,
\begin{equation}
R \subkjp \phr (T) = \eden \beta \phr \subkj (T) \, .
\end{equation}

The cross section for capture of a free electron decreases with
electron kinetic energy. Given the cross section, $\sigma \phr \subkjs
(E)$, for radiative recombination to shell $s$ forming ion $k,j$, the
radiative recombination rate coefficient for the shell, $\beta \phr
\subkjs (T)$, is obtained by integrating over all electron velocities,
analogous to Eq.~\ref{eq:alphacdicomp}, i.e., with no threshold energy
($I\subkjs \rightarrow 0$) and with $\sigma \cdi \subkjs (E)$ replaced
by $\sigma \phr \subkjs (E)$.  We have ignored radiation induced
recombination.  The total recombination rate coefficient is the sum
over all shells,
\begin{equation}
\beta \phr \subkj (T) =
\textstyle \sum \limits_{s=1} ^{N^{\hbox{\tiny s}}_{\hbox{\tiny k,j}}}
\beta \phr \subkjs (T) \, .
\end{equation}

For hydrogenic ions ($N_{e} = 1$) we use the formula originally
proposed by \citet{seaton59}, which is highly accurate
\citep{arnaud85,diffuseuniverse},
\begin{equation}
\beta \phr \subkj (T) = 
\beta_{_0} Z_{\hbox{\tiny k}} \lambda ^{\hbox{\tiny 1/2}}
\left[ 0.4288 + 0.5 \ln \lambda + 
        0.469\lambda^{\hbox{\tiny -1/3}} \right] \, ,
\end{equation}
where $\beta _{_0} = 5.197 \times 10^{-14} $ and where $\lambda =
Z_{\hbox{\tiny k}} ^{\hbox{\tiny 2}} (1.5789 \times 10^5/T)$.

The Atomic and Molecular Diagnostic Processes in Plasmas (AMDPP) group
has published fitting functions and parameters for radiative
recombination rate coefficients for many non-hydrogenic ions
\citep{badnell06}.  The functional form is
\begin{equation}
\beta \phr \subkj (T) = 
\frac{a \subkj}{\left( T/t^{\hbox{\tiny (0)}}\subkj \right) ^{\hbox{\tiny 1/2}}}
\frac
{ \left[ 1 + \left( T/t^{\hbox{\tiny (0)}}\subkj \right) ^{\hbox{\tiny 1/2}} \right] 
   ^{b' \subkj - 1} }
{ \left[ 1 + \left( T/t^{\hbox{\tiny (1)}}\subkj \right) ^{\hbox{\tiny 1/2}} \right] 
   ^{b' \subkj + 1} } \, ,
\end{equation}
where $a \subkj$, $t^{\hbox{\tiny (0)}}\subkj$, $t^{\hbox{\tiny
    (1)}}\subkj$ are tabulated fitting parameters, and $b' \subkj =
b\subkj + c\subkj \exp \{ -t^{\hbox{\tiny (2)}}\subkj/T \}$, where
$b\subkj$, $c\subkj$ and $t^{\hbox{\tiny (2)}}\subkj$ are additional
fitting parameters.  Data are tabulated for all elements from helium
to zinc for isoelectronic sequences up to the magnesium sequence
([Ne]3s$^{2}$, $N_{\rm e} = 12$).  The fitting functions neglect
narrow resonant spikes.

Since not all ionization stages have been tabulated by
\citet{badnell06}, we employed the fitting functions and parameters
for the simple power-law form published by \citet{arnaud85}, based
upon work of \citet{seaton59}, \citet{aldrovandi73}, and
\citet{shull82} for cases omitted from the AMDPP tables.  The
expression is
\begin{equation}
\beta \phr \subkj (T) = a \subkj T_{_4} ^{-b \subkj} \, ,
\end{equation}
where $T_{_4} = T/10^4$, and where $a \subkj$ and $b \subkj$ are the
fitting parameters tabulated by \citet{arnaud85} for all ions of
helium through nickel.

\subsubsection{Dielectronic Recombination}
\label{sec:die}

Dielectronic recombination often dominates over radiative
recombination.  In this process, a high energy free electron first
excites a bound deep inner shell electron prior to its capture in an
elevated excited state of the ion.  There are now two excited
electrons and an unfilled state in an inner shell.  Multiple channels
of relaxation for the ion are now available (of which one is also
auto-ionization).

In dielectronic recombination, the doubly excited ion works its way
back to the ground state via multiple radiative cascades.  At high
temperatures this process usually proceeds first by the decay of one
of the excited electrons to refill the empty inner shell by radiative
decay followed by a downward cascade of the remaining excited
electron.  At low temperatures, the dominant channel occurs when the
free electron is capture in a shell that is a resonant state to the
emptied inner shell.  The electron transitions rapidly and is then
followed by a downward cascade of the remaining excited electron.  The
reaction can be written
\begin{equation}
A \subkjp + {\rm e}_{\hbox{\tiny f}}^- 
\rightarrow A^{\ast \ast} \subkj 
\rightarrow A^{\ast} \subkj + \gamma 
\rightarrow A \subkjp + \textstyle \sum \gamma_i \, ,
\end{equation}
where the sum indicates that several recombination photons can be
emitted during the cascade process.

The dielectronic recombination rate for creation of ion $k,j$ due to
electron recombination with ion $k,j+1$ is obtained by multiplying the
total dielectronic recombination rate coefficient, $\beta \die \subkj
(T)$, by the electron number density,
\begin{equation}
R \subkjp \die (T) = \eden 
\beta \die \subkj (T) \, .
\end{equation}

Since the dominant channels for dielectronic recombination are
temperature dependent, the rate coefficient is double peaked.  For
this reason, previous fitting functions and parameters for the rate
coefficients were split into a low temperature regime
\citep{nuss83,nuss86,nuss87} and high temperature regime
\citep{aldrovandi73,shull82,arnaud85}.

A newer fitting function and accompanying parameter list for all
elements from helium to zinc and valid for temperatures ranging from
$T\simeq 100$ to $T\simeq 10^7$K has been made available by the AMDPP
group.  We used the fitting functions and parameters described in
\citet{altun07}, which are based upon a series of papers \citep[see
references in][]{badnell03, altun07}. The fitting function has the
form
\begin{equation}
\beta \die \subkj (T) = T^{\hbox{\tiny -3/2}} \, 
{\textstyle \sum \limits _{i=1}^{N\subkj}} 
c \subkji \exp \left\{ - \frac{t\subkji}{T} \right\} \, ,
\end{equation}
where $N \subkj$ is the number of fitting parameters for ion $k,j$,
and $c \subkji$ and $t\subkji$ are the fitting parameters. 



\subsection{Charge Exchange}
\label{sec:chargeexchange}

Charge exchange is the transfer of an electron from one ion to another
during a collision.  Since hydrogen is the most abundant species, a
charge exchange with a given metal ion $k,j$ is dominated either by
ionization ($k,j \rightarrow k,j+1$) from an ionized hydrogen (in
which the H$^{+}$ ion recombines with the exchanged
electron), or by recombination ($k,j-1 \rightarrow k,j$) via the
ionization of neutral hydrogen,
\begin{equation}
A \subkj + {\rm H}^{\hbox{\tiny +}} \leftrightarrow 
A \subkjp + {\rm H}^{\hbox{\tiny 0}}  \, .
\end{equation}
Helium is also relatively abundant and is the second most important
charge exchange channel,
\begin{equation}
A \subkjp + {\rm He}^{\hbox{\tiny 0}} \leftrightarrow 
A \subkj + {\rm He}^{\hbox{\tiny +}} \, .
\end{equation}

The rate for destruction of ion $k,j$ via charge exchange ionization 
with ionized hydrogen ($k=1,j=2$) is 
\begin{equation}
R ^{\hbox{\tiny xH$^+$}}_{\hbox{\tiny k,j}} (T) =
n_{\hbox{\tiny 1,2}} \alpha^{\hbox{\tiny xH$^+$}}_{\hbox{\tiny k,j}} (T) \, ,
\end{equation}
where $\alpha^{\hbox{\tiny xH$^+$}}_{\hbox{\tiny k,j}} (T)$ is the
ionization rate coefficient.  The rates for creation of ion $k,j-1$
via destruction of ion $k,j$ via charge exchange recombination with
neutral hydrogen ($k=1,j=1$) and with neutral helium ($k=2,j=1$) are
given by
\begin{equation}
\begin{array}{lcl}
\displaystyle
R ^{\hbox{\tiny xH}}_{\hbox{\tiny k,j}} (T) \!\!\!\! & = & \!\!\!\! 
n_{\hbox{\tiny 1,1}} \beta ^{\hbox{\tiny xH}}_{\hbox{\tiny k,j-1}} (T) 
 \\[10pt]
R ^{\hbox{\tiny xHe}}_{\hbox{\tiny k,j}} (T) \!\!\!\! & = & \!\!\!\! 
n_{\hbox{\tiny 2,1}} \beta ^{\hbox{\tiny xHe}}_{\hbox{\tiny k,j-1}} (T) \, ,
\end{array}
\end{equation}
where $ \beta ^{\hbox{\tiny xH}}_{\hbox{\tiny k,j-1}} (T) $ and
$\beta ^{\hbox{\tiny xHe}}_{\hbox{\tiny k,j-1}} (T)$ are the
respective recombination rate coefficients.

We computed the total charge exchange ionization and recombination
rate coefficients using the fitting function and parameters of
\citet{kingdon96}.  The rate coefficient for recombination to ion
$k,j$ via charge exchange from neutral hydrogen is given by
\begin{equation}
\beta \subkj ^{\hbox{\tiny xH}} (T) = 10^{-9}  
a\subkj T_{_4}^{b\subkj} \left[ 1 + c\subkj 
\exp \left\{ d\subkj T_{_4} \right\} \right] \, ,
\label{eq:exHion}
\end{equation}
where $T_{_4} = T/10^4$, and where $a\subkj$, $b\subkj$, $c\subkj$,
and $d\subkj$ are the fitting parameters.  The rate coefficient
for ionization of ion $k,j$ via charge exchange to neutral hydrogen
is obtained via detailed balancing,
\begin{equation}
\alpha ^{\hbox{\tiny xH$^+$}} \subkj (T) = 
\beta \subkj ^{\hbox{\tiny xH}} (T) \,
\exp \left\{ -\frac{\Delta E\subkj}{k T_{_4}} \right\} \, ,
\end{equation}
where the Boltzmann factor, $\Delta E\subkj /k$, is also tabulated by
\citet{kingdon96}.  

The computation of the recombination rate coefficient for charge
exchange with neutral helium, $ \beta \subkj^{\hbox{\tiny xHe}} (T)$,
is also obtained from Eq.~\ref{eq:exHion} using the parameters
applicable to these reactions.  The charge exchange ionization of
metals via ionized helium is not treated because, to date, there is
not a uniform set of published rates covering a wide range of
ions.

The fitting parameters provided by \citet{kingdon96} are presented for
all species up to and including zinc, but only for the first three
ionization stages.  For ions with $j\geq 4$, we use the asymptotic
formulae of \citet{ferland97},
\begin{equation}
\beta \subkj ^{\hbox{\tiny xH}} (T) = 1.92 \times 10^{-9} \, Z\subk \, ,
\end{equation}
for hydrogen, and
\begin{equation}
\beta \subkj ^{\hbox{\tiny xHe}} (T) = 5.4 \times 10^{-10} \, Z\subk \, ,
\end{equation}
for helium.  The different constants are due to the different reduced
masses of hydrogen and helium.

\subsection{Rate Equations}
\label{sec:rateeqs}

Here, we derive the rate equations, $dn \subkj /dt$, for all
ionization stages of all treatable species.  We assume two-level
atoms, effectively the ground state and the continuum\footnote{For a
brief discussion of the ramification of this assumption, see
Section~11.1 of {\it Hazy 2} \citep{hazy}.}.

For the following, we drop the explicit temperature dependence of all
rates and rate coefficients.  We remind the reader that for
recombination, the rate coefficients, $\beta \subkj$, are indexed to
the final state.  However, we use the convention that all rates, $R
\subkj$, are indexed by the initial state.

\subsubsection{Hydrogen}

Hydrogen is the simplest case because the channels for creation and
destruction involve only two adjacent ionization stages. Because of
this, the hydrogen rate equations for $n_{\hbox{\tiny 1,1}}$ and
$n_{\hbox{\tiny 1,2}}$ are antisymmetric,
\begin{equation}
\begin{array}{lcl}
\displaystyle
\frac{dn_{\hbox{\tiny 1,1}}}{dt} \!\!\!\! & = & \!\!\!\!  
n_{\hbox{\tiny 1,2}} \!
\left(
R^{\hbox{\tiny rec}}_{\hbox{\tiny 1,2}} +
R^{\hbox{\tiny xH$^{+}$}}_{\hbox{\tiny 1,2}}   
\right)
- n_{\hbox{\tiny 1,1}} \!
\left( 
R^{\hbox{\tiny ph}}_{\hbox{\tiny 1,1}} +  
R^{\hbox{\tiny coll}}_{\hbox{\tiny 1,1}} +  
R^{\hbox{\tiny xH}}_{\hbox{\tiny 1,1}}   
\right)  \\[12pt]
\displaystyle
\frac{dn_{\hbox{\tiny 1,2}}}{dt} \!\!\!\! & = & \!\!\!\!
\displaystyle - \frac{dn_{\hbox{\tiny 1,1}}}{dt}  \, .
\end{array}
\label{eq:hydrogenrateeq}
\end{equation}
The creation rates of $n_{\hbox{\tiny 1,1}}$ are due to the
recombination of $n_{\hbox{\tiny 1,2}}$ with free electrons,
$R^{\hbox{\tiny rec}}_{\hbox{\tiny 1,2}}$, and ionization charge
exchange from metals, $R^{\hbox{\tiny xH$^{+}$}}_{\hbox{\tiny 1,2}}$,
where
\begin{equation}
\begin{array}{lcl}
R^{\hbox{\tiny rec}}_{\hbox{\tiny 1,2}} \!\!\!\! & = & \!\!\!\! 
\eden \beta^{\hbox{\tiny phr}}_{\hbox{\tiny 1,1}} \\[8pt]
R^{\hbox{\tiny xH$^{+}$}}_{\hbox{\tiny 1,2}} \!\!\!\! & = & \!\!\!\!
\sum \limits_{\hbox{\tiny k=2}} \sum \limits_{\hbox{\tiny j=1}}^{\hbox{\tiny k}} 
    {\nkj} {\acth} \, .
\end{array}
\end{equation}
The destruction rates of $n_{\hbox{\tiny 1,1}}$ are due to
photoionization, $R^{\hbox{\tiny ph}}_{\hbox{\tiny 1,1}}$, collisional
ionization via free electrons, $R^{\hbox{\tiny coll}}_{\hbox{\tiny
1,1}}$, and recombination charge exchange to metals, $R^{\hbox{\tiny
xH}}_{\hbox{\tiny 1,1}}$, which ionizes neutral hydrogen, where
\begin{equation}
\begin{array}{lcl}
R^{\hbox{\tiny coll}}_{\hbox{\tiny 1,1}} \!\!\!\! & = & \!\!\!\!
\eden \alpha^{\hbox{\tiny cdi}}_{\hbox{\tiny 1,1}} \\[8pt]
R^{\hbox{\tiny xH}}_{\hbox{\tiny 1,1}} \!\!\!\! & = & \!\!\!\!  
\sum \limits_{\hbox{\tiny k=2}} \sum \limits_{\hbox{\tiny j=2}}^{\hbox{\tiny k+1}} 
    {\nkj} {\bcth} \, .
\end{array}
\end{equation}
Note that the negative of these rates are also the destruction and
creation rates of $n_{\hbox{\tiny 1,2}}$, respectively.

\subsubsection{Helium}

Helium has three ionization stages.  There is no published Auger
channel directly connecting the neutral and fully ionized stages;
however, non-zero dielectronic rate coefficients for the channel from
singly ionized to neutral helium exist.  The rate equation for neutral
helium is
\begin{equation}
\frac{dn_{\hbox{\tiny 2,1}}}{dt} = 
n_{\hbox{\tiny 2,2}} \!
\left(
  R^{\hbox{\tiny rec}}_{\hbox{\tiny 2,2}} 
+ R^{\hbox{\tiny xHe$^{+}$}}_{\hbox{\tiny 2,2}}   
\right)
- n_{\hbox{\tiny 2,1}} \!
\left( 
  R^{\hbox{\tiny ph}}_{\hbox{\tiny 2,1}}   
+ R^{\hbox{\tiny coll}}_{\hbox{\tiny 2,1}}   
+ R^{\hbox{\tiny xHe}}_{\hbox{\tiny 2,1}} 
\right) \, .
\label{eq:heliumIrateeq}
\end{equation}
The creation rates of $n_{\hbox{\tiny 2,1}}$ are due to the 
recombination channels of $n_{\hbox{\tiny 2,2}}$ with free electrons,
$R^{\hbox{\tiny rec}}_{\hbox{\tiny 2,2}}$, and ionization charge
exchange from metals, $R^{\hbox{\tiny xHe$^{+}$}}_{\hbox{\tiny 2,2}}$,
where
\begin{equation}
\begin{array}{lcl}
R^{\hbox{\tiny rec}}_{\hbox{\tiny 2,2}} \!\!\!\! & = & \!\!\!\! 
\eden \! \left( \beta^{\hbox{\tiny phr}}_{\hbox{\tiny 2,1}} +
\beta^{\hbox{\tiny die}}_{\hbox{\tiny 2,1}} \right)  \\[8pt]
R^{\hbox{\tiny xHe$^{+}$}}_{\hbox{\tiny 2,2}} \!\!\!\! & = & \!\!\!\!
n_{\hbox{\tiny 1,1}} \alpha ^{\hbox{\tiny xHe$^{+}$}}_{\hbox{\tiny 1,1}}
+ \sum \limits_{\hbox{\tiny k=3}} \sum \limits_{\hbox{\tiny j=1}}^{\hbox{\tiny k}} 
   {\nkj} \alpha ^{\hbox{\tiny xHe$^{+}$}} \subkj \, . 
\end{array}
\end{equation}
Recall, however, that we do not treat charge exchange ionization from
ionized helium, so $R^{\hbox{\tiny xHe$^{+}$}}_{\hbox{\tiny 2,2}}= 0$
for our work.  

The destruction rates of $n_{\hbox{\tiny 2,1}}$ are
due to photoionization, $R^{\hbox{\tiny ph}}_{\hbox{\tiny 2,1}}$,
collisional ionization via free electrons, $R^{\hbox{\tiny
coll}}_{\hbox{\tiny 2,1}}$, and recombination charge exchange to
metals, $R^{\hbox{\tiny xH}}_{\hbox{\tiny 2,1}}$, which singly
ionizes neutral helium, where
\begin{equation}
\begin{array}{lcl}
R^{\hbox{\tiny coll}}_{\hbox{\tiny 2,1}} \!\!\!\! & = & \!\!\!\!
\eden \alpha^{\hbox{\tiny cdi}}_{\hbox{\tiny 2,1}}   \\[8pt]
R^{\hbox{\tiny xHe}}_{\hbox{\tiny 2,1}}   \!\!\!\! & = & \!\!\!\!
n_{\hbox{\tiny 1,2}} \beta ^{\hbox{\tiny xHe}}_{\hbox{\tiny 1,1}}
+ \sum \limits_{\hbox{\tiny k=3}} \sum \limits_{\hbox{\tiny j=2}}^{\hbox{\tiny k+1}} 
   {\nkj} {\bcthe} \, ,
\end{array}
\end{equation}
respectively. For twice ionized helium, the rate equation
is
\begin{equation}
\frac{dn_{\hbox{\tiny 2,3}}}{dt} = 
n_{\hbox{\tiny 2,2}} \!
\left(
  R^{\hbox{\tiny ph}}_{\hbox{\tiny 2,2}} 
+ R^{\hbox{\tiny coll}}_{\hbox{\tiny 2,2}}  
+ R^{\hbox{\tiny xH$^+$}}_{\hbox{\tiny 2,2}} 
\right)
- n_{\hbox{\tiny 2,3}} \!
\left( 
R^{\hbox{\tiny rec}}_{\hbox{\tiny 2,3}} + 
R^{\hbox{\tiny xH}}_{\hbox{\tiny 2,3}}  
\right) \, .
\label{eq:heliumIIIrateeq}
\end{equation}
The creation rates via the destruction of $n_{\hbox{\tiny 2,2}}$ are
due to photoionization, $R^{\hbox{\tiny ph}}_{\hbox{\tiny 2,2}}$,
collisional ionization, $R^{\hbox{\tiny coll}}_{\hbox{\tiny 2,2}}$,
and ionization via charge exchange recombination to ionized hydrogen,
where
\begin{equation}
\begin{array}{lcl}
R^{\hbox{\tiny coll}}_{\hbox{\tiny 2,2}}  \!\!\!\! & = & \!\!\!\! 
  \eden \alpha^{\hbox{\tiny cdi}}_{\hbox{\tiny 2,2}}   \\[8pt]
R^{\hbox{\tiny xH$^+$}}_{\hbox{\tiny 2,2}}  \!\!\!\! & = & \!\!\!\! 
  n_{\hbox{\tiny 1,2}} \alpha^{\hbox{\tiny xH$^+$}}_{\hbox{\tiny 2,2}} \, .
\end{array}
\end{equation}
The destruction rates of $n_{\hbox{\tiny 2,3}}$ are due to the 
channels of recombination with free electrons, $R^{\hbox{\tiny
rec}}_{\hbox{\tiny 2,3}}$, and recombination via charge exchange
ionization of neutral hydrogen, where
\begin{equation}
\begin{array}{lcl}
R^{\hbox{\tiny rec}}_{\hbox{\tiny 2,3}} \!\!\!\! & = & \!\!\!\! 
   \eden \beta^{\hbox{\tiny phr}}_{\hbox{\tiny 2,2}} 
\\[8pt]
R^{\hbox{\tiny xH}}_{\hbox{\tiny 2,3}} \!\!\!\! & = & \!\!\!\!
   n_{\hbox{\tiny 1,1}} \beta^{\hbox{\tiny xH}}_{\hbox{\tiny 2,2}} \, .
\end{array}
\end{equation}
For singly ionized helium, the creation and destruction rates are
simply the negative of the sum of those for neutral and doubly ionized
helium,
\begin{equation}
\frac{dn_{\hbox{\tiny 2,2}}}{dt} = - \left(
 \frac{dn_{\hbox{\tiny 2,1}}}{dt}
+ \frac{dn_{\hbox{\tiny 2,3}}}{dt} \right) \, .
\label{eq:heliumIIrateeq}
\end{equation}

\subsubsection{Metals}

Here, we write out the rate equations for all ions with $k\geq 3$.
For what follows, let $R^{\hbox{\tiny ion}}_{\hbox{\tiny k,j-1}}$
denote the creation rate of $n_{\hbox{\tiny k,j}}$ via ionization of
$n_{\hbox{\tiny k,j-1}}$ and let $R^{\hbox{\tiny rec}}_{\hbox{\tiny
k,j+1}}$ denote the creation rate of $n_{\hbox{\tiny k,j}}$ via
recombination from initial state $n_{\hbox{\tiny k,j+1}}$.  Further,
let $R^{\hbox{\tiny rec}}_{\hbox{\tiny k,j}}$ denote the destruction
rate of $n_{\hbox{\tiny k,j}}$ via recombination to final state
$n_{\hbox{\tiny k,j-1}}$ and let $R^{\hbox{\tiny ion}}_{\hbox{\tiny
k,j}}$ denote the destruction of $n_{\hbox{\tiny k,j}}$ via ionization
to final state $n_{\hbox{\tiny k,j+1}}$.  We then write
\begin{equation}
\begin{array}{lcl}
{\displaystyle \frac{dn_{\hbox{\tiny k,j}}}{dt} } \!\! & = & \!\! 
n_{\hbox{\tiny k,j-1}} R^{\hbox{\tiny ion}}_{\hbox{\tiny k,j-1}}  
+  n_{\hbox{\tiny k,j+1}} R^{\hbox{\tiny rec}}_{\hbox{\tiny k,j+1}}  
+  {\textstyle \sum \limits_{\hbox{\tiny i=1}}^{\hbox{\tiny j-2}}} 
   n_{\hbox{\tiny k,i}} R^{\hbox{\tiny aug}}_{\hbox{\tiny k,i,j}} 
\\[12pt]
& & 
-  n_{\hbox{\tiny k,j}} \! \left( R^{\hbox{\tiny ion}}_{\hbox{\tiny k,j}}  
+  R^{\hbox{\tiny rec}}_{\hbox{\tiny k,j}}
+  {\textstyle \sum \limits_{\hbox{\tiny m=j+2}}^{\hbox{\tiny k-1}}} 
   R^{\hbox{\tiny aug}}_{\hbox{\tiny k,j,m}} \right) \, .
\end{array}
\label{eq:dnkjdt}
\end{equation}
The creation rate of $n \subkj$ via ionization destruction of adjacent
ion $n \subkjm$ is
\begin{equation}
R^{\hbox{\tiny ion}}_{\hbox{\tiny k,j-1}} 
=  R^{\hbox{\tiny ph}}_{\hbox{\tiny k,j-1}} 
+  \eden \! \left( \alpha^{\hbox{\tiny cdi}}_{\hbox{\tiny k,j-1}}
+                  \alpha^{\hbox{\tiny cea}}_{\hbox{\tiny k,j-1}} \right) 
+  n_{\hbox{\tiny 1,2}} \alpha^{\hbox{\tiny xH$^+$}}_{\hbox{\tiny k,j-1}}  \, .
\label{eq:Rionkjm}
\end{equation}
Note that we do not treat charge exchange ionization from ionized
helium.  The creation rate of $n \subkj$ via recombination destruction
of adjacent ion $n \subkjp$ is
\begin{equation}
R^{\hbox{\tiny rec}}_{\hbox{\tiny k,j+1}} 
=  \eden \! \left( \beta^{\hbox{\tiny phr}}_{\hbox{\tiny k,j}} 
+  \beta^{\hbox{\tiny die}}_{\hbox{\tiny k,j}} \right) 
+  n_{\hbox{\tiny 1,1}} \beta ^{\hbox{\tiny xH}}_{\hbox{\tiny k,j}}   
+  n_{\hbox{\tiny 2,1}} \beta ^{\hbox{\tiny xHe}}_{\hbox{\tiny k,j}}  \, .
\label{eq:Rreckjp}
\end{equation}

The recombination destruction rate of $n \subkj$ to adjacent stage $n
\subkjm$ and the ionization destruction rate of $n \subkj$ to adjacent
stage $n \subkjp$, are
\begin{equation}
\begin{array}{lcl}
R^{\hbox{\tiny rec}}_{\hbox{\tiny k,j}} \!\!\!\! & = & \!\!\!\! 
   \eden \! \left( \beta^{\hbox{\tiny phr}}_{\hbox{\tiny k,j-1}} 
+  \beta^{\hbox{\tiny die}}_{\hbox{\tiny k,j-1}} \right) 
+  n_{\hbox{\tiny 1,1}} \beta ^{\hbox{\tiny xH}}_{\hbox{\tiny k,j-1}}   
+  n_{\hbox{\tiny 2,1}} \beta ^{\hbox{\tiny xHe}}_{\hbox{\tiny k,j-1}}  
\\[10pt]
R^{\hbox{\tiny ion}}_{\hbox{\tiny k,j}} \!\!\!\! & = & \!\!\!\! 
   R^{\hbox{\tiny ph}}_{\hbox{\tiny k,j}} 
+  \eden \! \left( \alpha^{\hbox{\tiny cdi}}_{\hbox{\tiny k,j}} 
+                  \alpha^{\hbox{\tiny cea}}_{\hbox{\tiny k,j}} \right) 
+  n_{\hbox{\tiny 1,2}} \alpha^{\hbox{\tiny xH$^+$}}_{\hbox{\tiny k,j}}  \, ,
\end{array}
\label{eq:RionRreckj}
\end{equation}
respectively.  The summation terms in Eq.~\ref{eq:dnkjdt} account for
Auger ionization processes, which skip adjacent ionization stages.
All ions of species $k$ from the neutral stage to ionization stage $i
\leq j-2$ can contribute to the creation rate of $n_{\hbox{\tiny
k,j}}$ due to their destruction via Auger ionization
\begin{equation}
\frac{dn_{\hbox{\tiny k,j}}}{dt} \bigg|  \subaug 
= {\textstyle \sum \limits_{\hbox{\tiny i=1}}^{\hbox{\tiny j-2}} n_{\hbox{\tiny k,i}} }
  R^{\hbox{\tiny aug}}_{\hbox{\tiny k,i,j}} .
\label{eq:augkj-destruct} 
\end{equation}
Similarly, ion $k,j$ can be destroyed by Auger ionization to high
ionization final stage $m$, where $m \geq j + 2$,
\begin{equation}
\frac{dn_{\hbox{\tiny k,j}}}{dt} \bigg|  \subaug 
=  - \, n \subkj  {\textstyle \sum \limits_{\hbox{\tiny m=j+2}}^{\hbox{\tiny k-1}} }
   R^{\hbox{\tiny aug}}_{\hbox{\tiny k,j,m}} \, .
\label{eq:augkj-create}
\end{equation}
In practice, Auger ionization is a viable creation and destruction
channel only for $k \geq 4$.

\subsection{Equilibrium Solution}
\label{sec:solution}

If there are $N\subk$ atomic species included in the cloud model, then
there are $N = \sum_{\hbox{\tiny k}} (k+1) $ non-linear rate equations
to be solved, one for each $n\subkj$.  The equilibrium solution is
obtained when $dn_{\hbox{\tiny k,j}} / dt = 0 $ is satisfied for all
$k$ and $j$ (see Eqs.~\ref{eq:hydrogenrateeq}, \ref{eq:heliumIrateeq},
\ref{eq:heliumIIIrateeq}, \ref{eq:heliumIIrateeq}, and
\ref{eq:dnkjdt}).  The system of equations is ``closed'' by enforcing
charge density conservation, given by Eq.~\ref{eq:chargeconservation}.

The rate equations are non-linear because the collisional ionization
and recombination rates for ion $k,j$ include the product of the
electron density and the density of ion $k,j$, and the charge exchange
rates include the product of the number densities of the hydrogen and
helium ions and the density of ion $k,j$.

Here, we describe our method of linearizing the systems of equations.
We begin be rearranging the rate equations in terms of the ionization
fractions,
\begin{equation}
\displaystyle 
\frac{1}{f\subkj} \frac{df\subkj}{dt}  = 
\frac{1}{n\subkj} \frac{dn_{\hbox{\tiny k,j}}}{dt}  = 0 \, .
\label{eq:rearrange}
\end{equation}
As we show below, this formalism allows us to solve for the ratios of
the number densities of adjacent ionization stages $n\subkjp /
n\subkj$.  Defining $\Phi \subkj \equiv \Phi \subkj(\eden, T, J_{_E})
= n\subkjp / n\subkj $, the ionization fractions, $f \subkj$, are then
computed using a recursive formula.  Writing $f \subkj = P\subkj /
S\subk$, we have
\begin{equation}
P \subkj = P \subkjm \Phi \subkjm  \, ,
\qquad 
S \subk =  {\textstyle \sum \limits _{\hbox{\tiny j=1}}^{\hbox{\tiny k+1}} } P \subkj \, ,
\label{eq:defionfracs}
\end{equation}
where by definition $P_{\hbox{\tiny k,1}} = 1$.  Note that,
alternatively, $f_{\hbox{\tiny k,1}} = 1/S\subk$ and $f \subkj = \Phi
\subkjm f\subkjm$.  Thus, once all $\Phi \subkj$ are determined, all
ionization fractions are determined from which all ionic number
densities can be computed.

In order to linearize the equations, we adopt a method that reduces
the problem to solving for a single quantity, the electron density.
This requires that we decouple the hydrogen and helium from the metals
in order to remove the non-linearity arising from charge exchange with
metals.  To accomplish this, we first obtain an initial estimate for
the hydrogen, helium, and electron densities.  Using Brent's method,
we employ charge density conservation
(Eq.~\ref{eq:chargeconservation}) to solve for the equilibrium electron
density for a gas cloud composed of hydrogen and helium only.  For
hydrogen, we apply Eq.~\ref{eq:rearrange} and rearrange
Eq.~\ref{eq:hydrogenrateeq} to obtain,
\begin{equation}
\Phi_{\hbox{\tiny 1,1}} =  \displaystyle
 \frac{R\ph_{\hbox{\tiny 1,1}} + \eden \alpha^{\hbox{\tiny cdi}}_{\hbox{\tiny 1,1}}}
{\eden \beta^{\hbox{\tiny phr}}_{\hbox{\tiny 1,1}} } \, , \quad
f_{\hbox{\tiny 1,1}}  =  \displaystyle
 \frac{1}{1+\Phi_{\hbox{\tiny 1,1}}}\, , \quad
f_{\hbox{\tiny 2,1}}   =  \displaystyle \Phi_{\hbox{\tiny 1,1}} f_{\hbox{\tiny 1,1}} \, ,
\end{equation}
and for helium we rearrange Eqs.~\ref{eq:heliumIrateeq} and
\ref{eq:heliumIIrateeq}, to obtain
\begin{equation}
\begin{array}{lcllcl}
\Phi_{\hbox{\tiny 2,1}}  \!\!\! & = & \!\!\! \displaystyle
 \frac{R\ph_{\hbox{\tiny 2,1}} + \eden \alpha^{\hbox{\tiny cdi}}_{\hbox{\tiny 2,1}}}
{\eden \beta^{\hbox{\tiny phr}}_{\hbox{\tiny 2,1}} }  \, , &
\Phi_{\hbox{\tiny 2,2}}  \!\!\! & = & \!\!\! \displaystyle
 \frac{R\ph_{\hbox{\tiny 2,2}} + \eden \alpha^{\hbox{\tiny cdi}}_{\hbox{\tiny 2,2}}}
{\eden \beta^{\hbox{\tiny phr}}_{\hbox{\tiny 2,2}} } \, , \\[10pt]
f_{\hbox{\tiny 2,1}}  \!\!\! & = & \!\!\! \displaystyle
 \frac{1}{1+\Phi_{\hbox{\tiny 2,1}}+\Phi_{\hbox{\tiny 2,1}} \Phi_{\hbox{\tiny 2,2}}} \, , &
f_{\hbox{\tiny 2,2}}  \!\!\! & = & \!\!\! \displaystyle \Phi_{\hbox{\tiny 2,1}} f_{\hbox{\tiny 2,1}} \, , \,
f_{\hbox{\tiny 2,3}}   =  \displaystyle \Phi_{\hbox{\tiny 2,2}} f_{\hbox{\tiny 2,,2}} \, .
\end{array}
\end{equation}

With an initial estimate of the hydrogen, helium, and electron
densities, Eq.~\ref{eq:dnkjdt} for the metal ions can now be
rearranged for each $k,j$ to obtain the recursion formula
\begin{equation}
\Phi \subkj = \frac
{\displaystyle
R \supion \subkj + R\rec \subkj + R^{\, \hbox{\tiny A-out}} \subkj
- \Phi \supmo \subkjm R^{\, \hbox{\tiny ion}}_{\hbox{\tiny k,j-1}} 
- R^{\, \hbox{\tiny A-in}} \subkj
}
{\displaystyle 
R\rec \subkjp
} \, ,
\label{eq:phikj-metals}
\end{equation}
where $R^{\, \hbox{\tiny ion}}_{\hbox{\tiny k,j}}$ and $R\rec \subkj$
are given by Eq.~\ref{eq:RionRreckj}, $R^{\, \hbox{\tiny
    ion}}_{\hbox{\tiny k,j-1}}$ and $R\rec \subkjp$ by
Eqs.~\ref{eq:Rionkjm} and \ref{eq:Rreckjp}, respectively, and the Auger
destruction and creation rates are
\begin{equation}
R^{\, \hbox{\tiny A-out}} \subkj  =  \sum \limits _{\hbox{\tiny m=j+2}}^{\hbox{\tiny k-1}} R\aug _{\hbox{\tiny kjm}} \, , \quad
R^{\, \hbox{\tiny A-in}} \subkj   =  \displaystyle
 \sum \limits _{\hbox{\tiny i=1}}^{\hbox{\tiny j-2}} \left[ \prod _{\hbox{\tiny n=i}}^{\hbox{\tiny j-1}} \Phi \supmo _{\hbox{\tiny kn}} \right]
 R\aug _{\hbox{\tiny kij}} \, ,
\end{equation}
obtained from Eqs.~\ref{eq:augkj-destruct} and \ref{eq:augkj-create}.
For example, for $j=1$--4, we have 
\begin{equation}
\begin{array}{lcl}
\Phi_{\hbox{\tiny k,1}} \!\!\! & = & \!\!\!  R^{\, \prime}_{\hbox{\tiny k,1}} / R\rec_{\hbox{\tiny k,2}} \, ,  \\[7pt]
\Phi_{\hbox{\tiny k,2}}  \!\!\! & = & \!\!\!  
\left(  R^{\, \prime}_{\hbox{\tiny k,2}}
- \Phi \supmo _{\hbox{\tiny k,1}} R^{\, \hbox{\tiny ion}}_{\hbox{\tiny k,1}} \right) 
/ R\rec_{\hbox{\tiny k,3}} \, , \\[7pt]
\Phi_{\hbox{\tiny k,3}}  \!\!\! & = & \!\!\! 
\left(  R^{\, \prime}_{\hbox{\tiny k,3}}
- \Phi \supmo _{\hbox{\tiny k,2}} R^{\, \hbox{\tiny ion}}_{\hbox{\tiny k,2}} 
- \Phi \supmo _{\hbox{\tiny k,1}}  \Phi \supmo _{\hbox{\tiny k,2}} R^{\, \hbox{\tiny aug}}_{\hbox{\tiny k,1,3}} \right) 
/ R\rec_{\hbox{\tiny k,4}} \, , \\[7pt]
\Phi_{\hbox{\tiny k,4}}  \!\!\! & = & \!\!\! 
\left( R^{\, \prime}_{\hbox{\tiny k,4}}
- \Phi \supmo _{\hbox{\tiny k,3}} R^{\, \hbox{\tiny ion}}_{\hbox{\tiny k,3}} 
-  \Phi \supmo _{\hbox{\tiny k,1}} \Phi \supmo _{\hbox{\tiny k,2}} \Phi \supmo _{\hbox{\tiny k,3}} R^{\, \hbox{\tiny aug}}_{\hbox{\tiny k,1,4}} 
- \Phi \supmo _{\hbox{\tiny k,2}}  \Phi \supmo _{\hbox{\tiny k,3}} R^{\, \hbox{\tiny aug}}_{\hbox{\tiny k,2,4}} \right) 
/ R\rec_{\hbox{\tiny k,5}} \, .
\end{array}
\end{equation}
where we combined the destruction rates into the single term $R^{\,
  \prime} \subkj = R \supion \subkj + R\rec \subkj + R^{\, \hbox{\tiny
    A-out}}_{\hbox{\tiny k,j}} $.  From the $\Phi_{\hbox{\tiny k,j}}$,
we apply Eq.~\ref{eq:defionfracs} to compute the ionization fractions
for all metal ions.  We then refine the hydrogen and helium ionization
fractions by including charge exchange with metals ions,
\begin{equation}
\begin{array}{lcl}
\Phi_{\hbox{\tiny 1,1}} \!\!\! & = & \!\!\!  \displaystyle
\frac{\displaystyle R\ph_{\hbox{\tiny 1,1}} + \eden \alpha^{\hbox{\tiny cdi}}_{\hbox{\tiny 1,1}} 
 + n_{\hbox{\tiny A}} \textstyle \sum \limits_{\hbox{\tiny k=2}} \eta\subk \sum \limits_{\hbox{\tiny j=2}}^{\hbox{\tiny k+1}} f\subkj {\bcth} }  
{\displaystyle \eden \beta^{\hbox{\tiny phr}}_{\hbox{\tiny 1,1}} 
 + n_{\hbox{\tiny A}} \textstyle \sum \limits_{\hbox{\tiny k=2}} \eta\subk \sum \limits_{\hbox{\tiny j=1}}^{\hbox{\tiny k}} f\subkj {\acth} }  \, , \\[15pt]
\Phi_{\hbox{\tiny 2,1}}  \!\!\! & = & \!\!\! \displaystyle
 \frac{\displaystyle R\ph_{\hbox{\tiny 2,1}} + \eden \alpha^{\hbox{\tiny cdi}}_{\hbox{\tiny 2,1}} 
+ n_{\hbox{\tiny A}} \eta _{\hbox{\tiny 1}} f_{\hbox{\tiny 1,2}} \beta ^{\hbox{\tiny xHe}}_{\hbox{\tiny 1,1}}
+ n_{\hbox{\tiny A}} \textstyle \sum \limits_{\hbox{\tiny k=3}} \eta\subk \sum \limits_{\hbox{\tiny j=2}}^{\hbox{\tiny k+1}} f\subkj {\bcthe} }
{\displaystyle \eden \beta^{\hbox{\tiny phr}}_{\hbox{\tiny 2,1}} }  \, , \\[15pt]
\Phi_{\hbox{\tiny 2,2}}  \!\!\! & = & \!\!\! \displaystyle
 \frac{\displaystyle R\ph_{\hbox{\tiny 2,2}} + \eden \alpha^{\hbox{\tiny cdi}}_{\hbox{\tiny 2,2}} 
 + n_{\hbox{\tiny A}} \eta _{\hbox{\tiny 1}} f_{\hbox{\tiny 1,2}} \alpha^{\hbox{\tiny xH$^+$}}_{\hbox{\tiny 2,2}}  }
{\displaystyle \eden \beta^{\hbox{\tiny phr}}_{\hbox{\tiny 2,2}} + n_{\hbox{\tiny A}} \eta _{\hbox{\tiny 1}} f_{\hbox{\tiny 1,1}} \beta^{\hbox{\tiny xH}}_{\hbox{\tiny 2,2}} } \, ,
\end{array}
\label{eq:HHerefine}
\end{equation}

\begin{figure*}[hbt]
\epsscale{1.2}
\plotone{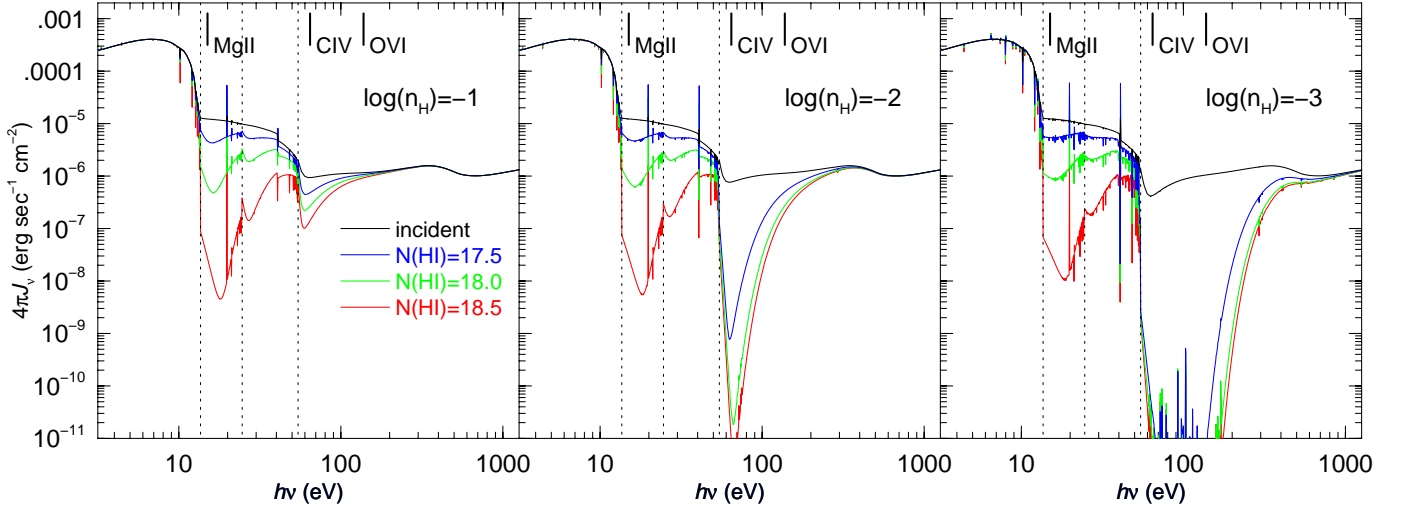}
\caption{The mean intensity of the attenuated ionizing spectrum
  transmitted though various cloud models as a function of photon
  energy [eV].  The black curves are the incident UVB.  Blue, green
  and, red curves are the attenuated spectrum after having
  passed through cloud models with fixed $\log N\subHnot= 17.5$, 18.0,
  and 18.5, respectively.  Vertical dotted lines give the ground-state
  ionization edges of H$^{\hbox{\tiny 0}}$ ({\HI}, 13.6 eV) ,
  He$^{\hbox{\tiny 0}}$ ({\HeI}, 24.6 eV), and He$^{+}$ ({\HeII}, 54.4
  eV).  For reference, the ground-state ionization edges of Mg$^{+}$
  ({\MgII}), C$^{\hbox{\tiny +3}}$ ({\CIV}), and O$^{\hbox{\tiny +5}}$
  ({\OVI}) are shown as vertical ticks.  (left) Results for cloud
  hydrogen number density $\log n\subH = -1$.  (center) Results for
  $\log n\subH = -2$.  (right) Results for $\log n\subH = -3$.  As
  $N({\HI}) \equiv N\subHnot$ of a cloud model with fixed $n\subH$
  increases, the physical depth of the cloud increases and the softer
  the ionizing spectrum becomes as more photons are absorbed due to
  the ionization of hydrogen and helium.  This alters the ionization
  balance of metals such as Mg$^{+}$, C$^{\hbox{\tiny +3}}$, and
  O$^{\hbox{\tiny +5}}$ as a function of depth into the cloud.  For
  fixed $N({\HI})$, model clouds with lower $n\subH$ have greater
  physical depth than higher $n\subH$ model clouds, resulting in much
  greater attenuation due to He$^{+}$ ionization relative to the
  hydrogen ionization edge.}
\label{fig:shield}
\end{figure*}

We remind the reader that we do not treat charge exchange ionization
of metals from ionized helium (which would appear as a additional
recombination term in the denominator of the expression for
$\Phi_{\hbox{\tiny 2,1}}$).

Using Brent's method, we iteratively apply
Eqs.~\ref{eq:phikj-metals}--\ref{eq:HHerefine} to converge on the full
equilibrium solution by enforcing charge density conservation via
Eq.~\ref{eq:chargeconservation}.  The method solves for the logarithm
of the equilibrium electron density to a precision of $1\times
10^{-20}$. The high precision is required in order to constrain ions
that yield small donations to the free electron pool; mostly these are
the elements with the lowest abundances.  Since the initial estimate
of the hydrogen, helium, and electron density from a zero metallicity
gas typically provides $\log \eden$ to 2-3 decimal points of accuracy,
usually only 5-7 iterations are required to converge $\log \eden$ to
20 decimal points of accuracy.

In the case of low ionization clouds, the ions with the least
constrained number densities are the high ionization stages of the low
abundance species, since they contribute negligibly to the free
electron pool.  In the case of high ionization clouds, the same
applies to the low ionization stages of the low abundance species.  In
other words, the method's strength is that it best constrains the
number densities of the ionic stages that contribute the most to the
electron pool.  Once all ionization fractions are solved, the ion
number densities are computed from $n\subkj = f\subkj n\subk = f\subkj
\eta \subk n_{\hbox{\tiny A}}$.

\section{The Optically Thin Constraint}
\label{sec:thingas}

Because we do not yet treat radiative transfer\footnote{We are
  currently implementing and testing self shielding in the code, which
  we will present in a second paper in this series.}  through the
cloud models (grid cells), currently {\code} is appropriate only for
optically thin gas \citep[also see][]{verner90}.  By optically thin,
we employ the definition that the optical depth is less than unity at
the hydrogen ionization edge (13.6 eV), and at both the neutral and
singly ionized helium ionization edges (24.6 and 54.4 eV,
respectively), which dominate modification of the ionizing SED.

To illustrate how the ionization edges modify the ionizing SED as
cloud models become progressively more optically thick, we plot the
mean intensity of the attenuated SED transmitted though various Cloudy
13.03 models as a function of photon energy in
Figure~\ref{fig:shield}.  The black curves are the incident
\citet{haardt11} UVB for $z=0$.  The clouds models have metallicity of
0.1 solar.  Blue, green and, red curves show the attenuated
transmitted spectrum after having passed through cloud models with
fixed $\log N({\HI}) \equiv \log N\subHnot = 17.5$, 18.0, and 18.5,
respectively.  Three hydrogen densities are illustrated, $\log n\subH
= -1$, $-2$, and $-3$.  The ground-state ionization edges of
H$^{\hbox{\tiny 0}}$, He$^{\hbox{\tiny 0}}$, and He$^{+}$ are shown as
vertical dotted lines.  For reference, vertical ticks indicate the
ground-state ionization edges of Mg$^{+}$ ({\MgII}), C$^{\hbox{\tiny
    +3}}$ ({\CIV}), and O$^{\hbox{\tiny +5}}$ ({\OVI}).  Clearly, the
number density of ionizing photons for these important metal species
can be substantially reduced with depth into the cloud model,
resulting in ionization structure in the cloud model and lower
ionization conditions in the shielded regions.

\begin{figure}[thb]
\epsscale{0.9}
\plotone{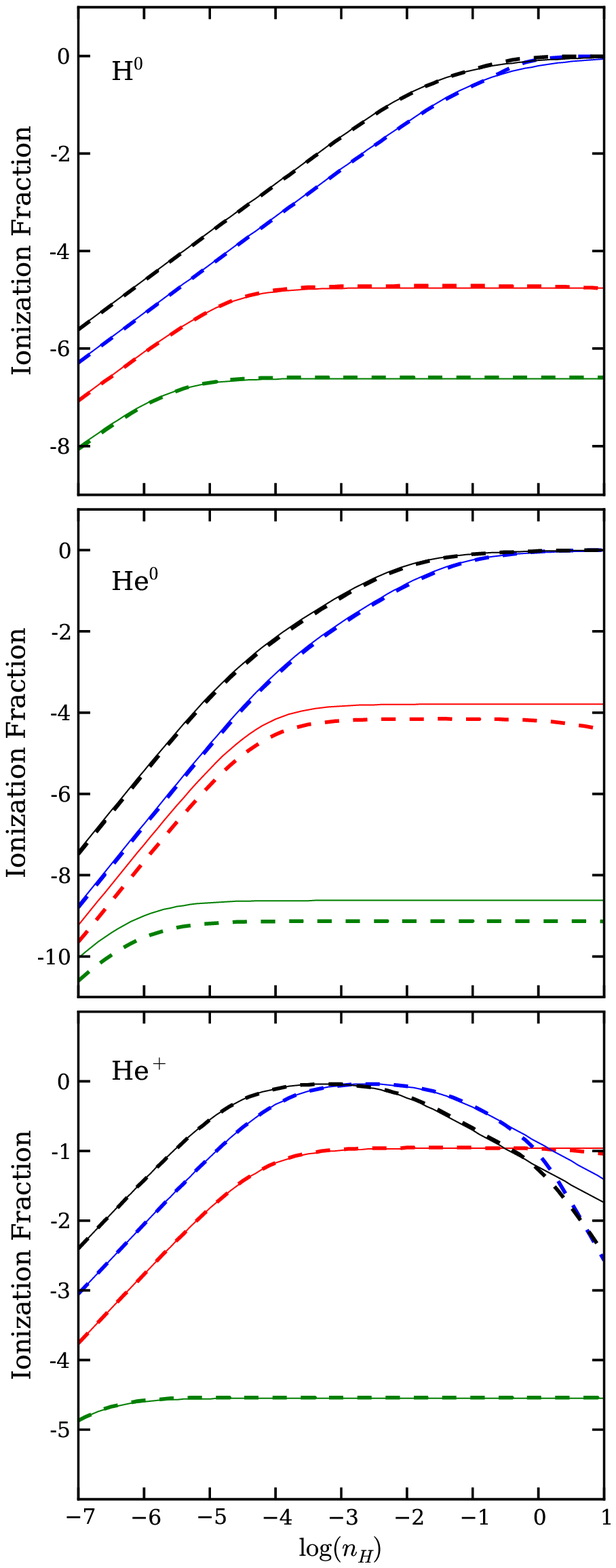}
\caption{The logarithm of the ionization fractions for H$^{\hbox{\tiny
      0}}$ (top), He$^{\hbox{\tiny 0}}$ (center), and He$^{+}$
      (bottom) as a function of hydrogen number density, $\log n\subH$
      for optically thin cloud models.  Dashed curves are the results
      from Cloudy and solid curves are the results from {\code}.
      Four temperatures are shown, $\log T= 3$ (black), $\log T= 4$
      (blue), $\log T= 5$ (red), $\log T= 6$ (green).  Our code
      {\code} is in excellent agreement with Cloudy over a large
      range of $n\subH$ and $T$, with no more than a factor of 2--3
      discrepancies for He$^{\hbox{\tiny 0}}$ at $\log T \geq 5$ and a
      divergence for He$^{+}$ at $\log T \leq 4$ for $\log n\subH > 0$.}
\label{fig:fh1}
\end{figure}

Since the optical depth is the product of the cross section for
bound-free absorption and the column density of the absorbing ion, we
can determine the upper limit on the H$^{\hbox{\tiny 0}}$,
He$^{\hbox{\tiny 0}}$, and He$^{+}$ column densities that satisfy our
criterion of an upper limit of unity optical depth at the respective
ionization edges. For ground state hydrogen and singly ionized helium,
the optical depth at the ionization edge is \citep{menzel35},
\begin{equation}
\tau = N \sigma = 6.304 \times 10^{-18} \cdot \frac{N}{Z^{2}} 
\left( \frac{\mu}{m_e} \right) ^{-1}  \, ,
\end{equation}
where $N$ is the column density, $\sigma$ is the bound-free cross
section for at the ionization energy from the ground state, $Z$ is the
number or protons in the nucleus, and $\mu$ is the reduced mass of the
electron, $\mu = m_e/(1+m_e/M\subk)$, where $M\subk$ is the nuclear mass of
species $k$.  For hydrogen, $Z=1$ and $\mu/m_e = 0.99946$, and for
helium, $Z=2$ and $\mu/m_e = 0.99986$.  Thus, for $\tau \leq 1$, the
cloud model is constrained to have $N\subHnot \leq 1.58\times 10^{17}$
cm$^{-2}$ for neutral hydrogen and $N\subHeplus \leq 6.34 \times
10^{17}$ cm$^{-2}$ for singly ionized helium.

For ground state neutral helium, the optical depth of the ionization
edge is \citep{vardya64},
\begin{equation}
\tau = N \sigma = 1.339 \times 10^{-18} \cdot N Z^{4}_{\rm eff} 
\left( \frac{\mu}{m_e} \right) ^{2} \lambda^3 g_{\hbox{\tiny II}} \, ,
\end{equation}
where $Z_{\rm eff} = 1.3343$ is the effective nuclear charge due to
screening, $\lambda = 504.19$~{\AA} is the wavelength at the
ionization edge, and $g_{\hbox{\tiny II}} \simeq 0.827$ is the
bound-free Gaunt factor \citep{menzel35}.  Evaluating, we obtain
the constraint
$N\subHenot \leq 2.77 \times 10^{17}$ cm$^{-2}$ for $\tau \leq 1$.

In terms of the limiting column densities, the optically thin
constraints place upper limits on the model cloud depth, $L_{\rm
  max}$, which is to say, in our application, that it places an upper
limit on the grid cell size before self-shielding must be treated.
Geometrically, $L\subHnot = N\subHnot/(f\subHnot n\subH)$, where
$f\subHnot = n\subHnot/n\subH$ is the ionization fraction of neutral
hydrogen.  When $N\subHnot = 1.58\times 10^{17}$ cm$^{-2}$, then
$L\subHnot$ corresponds to the cloud depth at which the optical depth
at the hydrogen ionization edge is unity.  Similarly, for neutral
helium, the upper limit is $L\subHenot = N\subHenot/(f\subHenot n\subH
\cdot \eta \subHe/\eta \subH)$ for $N\subHenot = 2.77 \times 10^{17}$
cm$^{-2}$ , where $\eta \subH$ and $\eta \subHe$ are the abundance
fractions of hydrogen and helium, respectively (see
Section~\ref{sec:abundances}).  For singly ionized helium,
$L\subHeplus = N\subHeplus / (f\subHeplus n\subH \cdot \eta
\subHe/\eta \subH)$ for $N\subHeplus = 6.34 \times 10^{17}$ cm$^{-2}$.


Assuming a fairly constant ratio $\eta\subHe / \eta\subH \simeq 0.1$,
we can rewrite the cell upper limits as follows,
\begin{equation}
\begin{array}{rcl}
L\subHnot^{\rm max} & \simeq & 0.5 \cdot ({0.01}/{f\subHnot}) 
({0.01}/{n\subH})  \\[5pt]
L\subHenot^{\rm max} & \simeq & 9 \cdot
({0.01}/{f\subHenot}) ({0.01}/{n\subH}) \quad {\rm kpc} \\[5pt]
L\subHeplus^{\rm max} & \simeq & 20 \cdot ({0.01}/{f\subHeplus}) 
({0.01}/{n\subH})  \, .
\end{array}
\label{eq:maxcell}
\end{equation}
For the criterion of optically thin radiative transfer, the upper
limit on the cloud depth, $L_{\rm max}$, is the minimum of
$L\subHnot$, $L\subHenot$, and $L\subHeplus$.  As the radiation
propagates deeper into the cloud, the mean intensity of the SED will
be modified by ionization (the SED is ``softened'' by the removal of
high energy photons).  As such, $L_{\rm max}$ is the depth into the
cloud to which the ionization structure is constant.

The minimum comoving cell size for the hydroART simulations is roughly
$30~h^{-1}$ pc ($0.03~h^{-1}$ kpc) and the proper size decreases with
redshift in proportion to $1/(1+z)$.  From Eq.~\ref{eq:maxcell}, we
see that only in cases where the product of the ionization fraction
and the hydrogen number density exceed $10^{-4}$ does the maximum cell
size decrease from the fiducial values of 0.5, 9 and 20 kpc for the
respective ionization edges.  In Figure~\ref{fig:fh1}, we plot the
ionization fractions for H$^{\hbox{\tiny 0}}$ (top), He$^{\hbox{\tiny
    0}}$ (center), and He$^{+}$ (bottom) as a function of hydrogen
number density, $\log n\subH$, for constant density and isothermal
optically thin clouds.  Four temperatures are presented, $\log T = 3$,
4, 5, and 6 as black, blue, red, and green curves, respectively.
Examining the behavior of the ionization fractions and propagating
them through Eq.~\ref{eq:maxcell}, we find that {\code} is currently
not valid for ``cold'' cells ($\log T < 4$) with densities $\log
n\subH > -2$ nor for ``warm/hot'' cells ($\log T \simeq 5$) with
densities $\log n\subH > -1$.
 
\section{Comparison to Cloudy}
\label{sec:compare}

\begin{figure*}[thb]
\epsscale{1.20}
\plotone{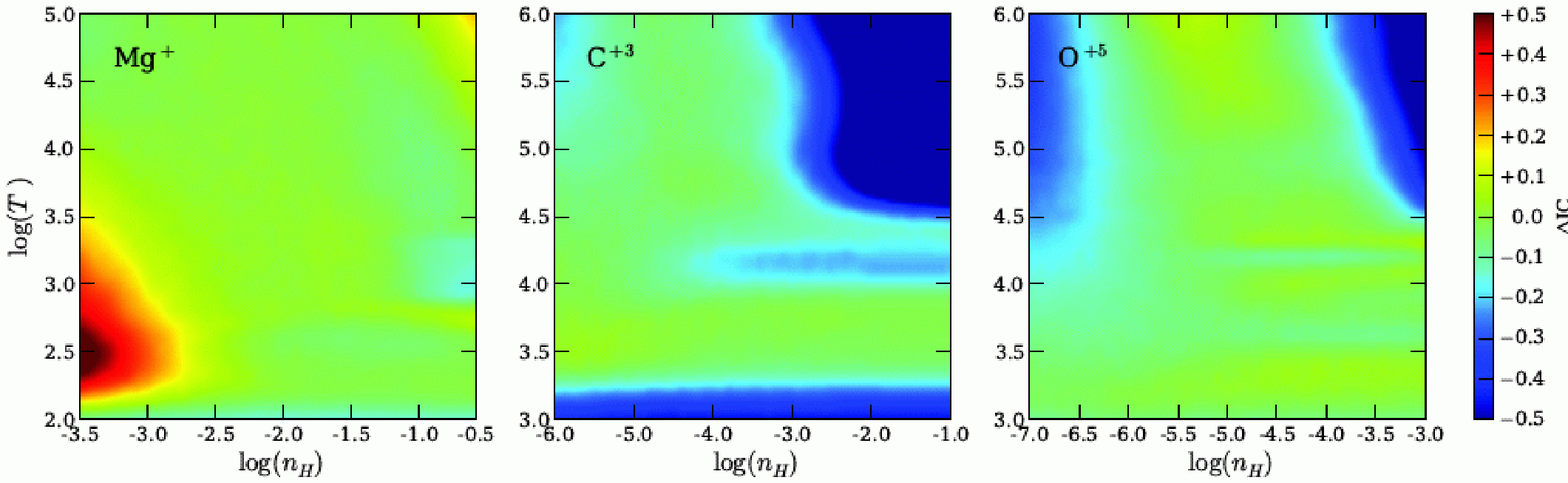}
\caption{A comparison of the ionization corrections, ${\rm IC} = \log
  \left\{ \fXj/\fHnot \right\}$ between {\code} and Cloudy.  The
  logarithmic difference if the ionization corrections, $\Delta {\rm
    IC}$, is plotted as a function of hydrogen number density and
  temperature of the cloud model. $\Delta {\rm IC} = 0$ indicates full
  agreement between the models.  (left) $\Delta {\rm IC}$ for ${\rm
    X}^{\hbox{\tiny j}} = {\rm Mg}^{+}$.  (center) $\Delta {\rm IC}$
  for C$^{\hbox{\tiny +3}}$.  (right) $\Delta {\rm IC}$ for
  O$^{\hbox{\tiny +5}}$.  The green shading represents $\Delta {\rm
    IC} \simeq \pm 0.05$, where agreement between the two codes is
  within the typical observational uncertainties of measured column
  densities.  Blue and red areas are regions where the ionization
  fractions of the metal ions are vanishingly small and are, for the
  most part, an artifact of the manner in which such ionization
  fractions are output by Cloudy (see text).}
\label{fig:amber}
\end{figure*}

The most important quantities for comparing ionization codes are the
ionization fractions.  In particular, those of hydrogen and helium are
critical since they dictate the conditions of the onset of self
shielding to the ionizing radiation.  To compare these ionization
fractions between {\code} and Cloudy 13.03, we ran both codes and
created a grid of cloud models while enforcing the optically thin
constraint.  We present the ionization fractions in
Figure~\ref{fig:fh1}. The dashed curves are Cloudy models and the
solid curves are {\code} models.  For both codes, we assume constant
density isothermal cloud models with no dust, no cosmic ray heating, a
metallicity of 10\% solar, and a \citet{haardt11} UVB at $z=0$.

{\code} is in general excellent agreement with Cloudy.  However, there
are deviations of up to a factor of two in $\fHenot$ for $\log T \geq
5$ at all $n\subH$ and for $\fHeplus$ for $\log T \simeq 3$ at $\log
n\subH > 0$. At these temperature, collisional ionization is becoming
more important relative to photoionization.  We have not identified
the source of the discrepancy with the He$^0$ ion in this regime.  One
main difference between {\code} and Cloudy is that Cloudy handles
recombination levels, whereas {\code} assumes two-level ions (ground
state and the continuum). In Cloudy, the He$^0$ ion includes the full
helium isoelectronic sequence \citep{porter05}.


We now compare the ionization fractions between {\code} and Cloudy for
the metal ions Mg$^{+}$, C$^{\hbox{\tiny +3}}$, and O$^{\hbox{\tiny
    +5}}$.  For this discussion, we slightly modify our notation from
Section~\ref{sec:themodel} such that species $k$ in ionization stage
$j$ is denoted by the somewhat more familiar notation ${\rm
  X}^{\hbox{\tiny j}}$ denoting species ${\rm X}$ in ionization stage
$j$.

For a constant density gas cloud, the column density of species ${\rm
  X}$ in ionization stage $j$, denoted $\NXj$, is
\begin{equation}
\NXj = \nXj L = \fXj n_{\hbox{\tiny X}} L, 
\label{eq:modelNxj}
\end{equation}
where the number density of ion ${\rm X}^{\hbox{\tiny j}}$ is $\nXj$,
the ionization fraction (from the ionization model) is $\fXj$, and the
number density of species ${\rm X}$ is $n_{\hbox{\tiny X}} = \left(
\eta_{\hbox{\tiny X}} / \eta \subH \right) n\subH$.  The quantity $L$
is the cloud depth.  Observers often combine the measured {\HI} column
density, $N\subHI$, with ionization modeling to estimate the
pathlength through the absorbing gas, using the expression $L =
N\subHI/(\fHnot n\subH)$.  For our comparison with Cloudy, we adopt
this convention because, observationally, $N\subHI$ is a directly
measurable quantity, whereas $L$ is not and must be inferred from the
ionization modeling.

Substituting the above expression $L$ into Eq.~\ref{eq:modelNxj}, we
obtain the ``observer's''expression for the cloud column density for
metal ion ${\rm X}^{\hbox{\tiny j}}$,
\begin{equation}
\log \NXj = \log N\subHI + \log \left\{
\frac{\fXj}{\fHnot} \right\} + \log \left\{
\frac{n_{\hbox{\tiny X}}}{n\subH} \right\}  \, ,
\label{eq:NXj}
\end{equation}
where $n\subX/n\subH = \left( \eta_{\hbox{\tiny X}} / \eta \subH
\right)$ corresponds to the abundance ratio $\left( {\rm X} / {\rm H}
\right)$.

The second term on the right hand side of Equation~\ref{eq:NXj} is
known as the ionization correction, $ {\rm IC} = \log \left\{
\fXj/\fHnot \right\}$.  This term is the critical quantity from
ionization models that, for observational work, allows the {\it
  measured\/} column densities to be used as constraints for inferring
gas phase abundances by solving $\log (\hbox{X/H}) = \log \NXj - \log
N\subHI - {\rm IC}$.  However, our application with the simulations
will be slightly different \citep[see][]{cwc-qals14}; we know the
line-of-sight pathlength through the cell and the elemental abundances
in the grid cell in the simulation and employ {\code} to determine the
cell column densities using Eq.~\ref{eq:modelNxj}.

Given the methods of application described above, we focus on the
ionization correction as the central quantity for comparing 
{\code} and Cloudy.  We define
\begin{equation}
\Delta {\rm IC} = \log \left\{ \frac{\fXj}{\fHnot} \right\} 
-  \log \left\{ \frac{\fXj}{\fHnot} \right\} _{\hbox{\tiny Cloudy}} \, ,
\label{eq:deltaIC}
\end{equation}
which, for fixed $N\subHI$ and $n\subX/n\subH$, provides a direct
measure of logarithmic difference in the calculated column density
between the two ionization models,
\begin{equation}
\left (\log \NXj \right) _{\hbox{\tiny HARTRATE}} 
- \left( \log \NXj \right) _{\hbox{\tiny Cloudy}} 
= \Delta {\rm IC} \, .  
\end{equation}
We computed $\Delta {\rm IC}$ as a function of $n\subH$ and
$T$ over the range $-7 \leq \log n\subH \leq 0 $ and $2 \leq \log T
\leq 7 $ for constant density isothermal cloud models.  We apply
Eq.~\ref{eq:maxcell} to ensure that the clouds are optically thin at
the hydrogen and helium ionization edges (which means $N\subHI$ varies
with $n\subH$).  For both {\code} and Cloudy, we assume no dust, no
cosmic ray heating, a metallicity of 10\% solar, and a
\citet{haardt01,haardt11} UVB at $z=0$ for the ionizing spectrum.

In Figure~\ref{fig:amber}, we plot smoothed $\Delta {\rm IC}$ surfaces
as a function of $\log n\subH$ and $\log T$ for the three commonly
observed ions, ${\rm Mg}^{+}$ (left panel), C$^{\hbox{\tiny +3}}$
(center), and O$^{\hbox{\tiny +5}}$ (right).  The presented ranges of
the density and temperature vary slightly from ion to ion based on the
appropriate ranges of gas phase in which they are dominant
\citep[see][]{cwc-qals14}.  The values of $\Delta {\rm IC}$ are
provided in the color bar.

Since typical uncertainties in observed column density measurements
are $\delta \log N \simeq \pm 0.05$ in the logarithm, values of
$\Delta {\rm IC} \leq \pm 0.05$ between the two ionization models
would be consistent with typical measurement errors in $\log \NXj$.
This level of uncertainty corresponds to the green area of the
surface in Figure~\ref{fig:amber}.  Thus, the green area provides the
$\log n\subH$--$T$ ranges over which the difference in the ionization
correction between {\code} and Cloudy are within reasonable
observational measurement uncertainties, and can therefore be
considered to yield column densities that are consistent within
practical errors.

The larger departures in $\Delta {\rm IC}$ (blue and red regions) are
due to (1) the linearization method employed for {\code} (see
Section~\ref{sec:solution}), and (2) the convention within Cloudy to
fix $\log \fXj = -30$ or $-50$ in the cases of very small ionization
fractions.  For (1), it is because we converge the rate matrix using
charge density conservation, per Eq.~\ref{eq:chargeconservation}, so
that ions contributing negligibly to the free electron density do not
have robustly constrained ionization fractions.  Since, in these
cases, the ionization fractions are typically on the order of
$10^{-20}$ or lower (recall that we conserve charge to this tolerance
level), these regions of gas phase space are not abundant in the ion.
As such, these phase space regions do not contribute to absorption
lines from the ion, and since one of our main goals is to study the
absorption properties of the gas in the simulations
\citep[e.g.,][]{cwc-qals14}, the absolute accuracy in the ionization
correction for these ions does not impact our science goals.

\section{Conclusions}
\label{sec:conclude}

We have presented the code {\code} for computing the equilibrium
ionization conditions for astrophysical gaseous environments.  The
main motivation for developing {\code} is to apply it to AMR
cosmological simulation in order to study the chemical and ionization
conditions of the circumgalactic medium in simulated galaxies using
the absorption line technique.  For a first application of the code to
AMR cosmological simulations, see \citet{cwc-qals14}.  A stand-alone
version of {\code} also exists, and has been applied to observational
data \citep[see][]{cwc-q1317, ggk-q1317}.

The physical gas processes included in {\code} are photoionization,
Auger ionization, direct collisional ionization, excitation
auto-ionization, charge exchange ionization, radiative recombination,
dielectronic recombination, and charge exchange recombination.
Currently, the code handles only optically thin gas.  Treatment of
optically thick gas will be presented in a companion paper.

{\code} is designed to take a minimum number of inputs to define a
``cloud'' model. The inputs are the gas hydrogen density, $n\subH$,
equilibrium temperature, $T$, and the mass fractions of all atomic
species.  To define the ionizing spectrum, the required inputs
are the redshift, $z$, for the \citet{haardt11} UVB, and if desired,
the masses, ages, metallicities, and locations of stellar populations
for the Starburst99 \citep{Sb99-99} SED models.

We compared {\code} to Cloudy 13.03 by examining the ionization
fractions of neutral hydrogen, neutral helium, and singly ionized
helium.  In the optically thin regime, the ionization fractions are
highly consistent, except for a factor of 2--3 difference in the
neutral helium values for $\log T \geq 5$.  We also presented a
comparison of the ionization corrections for the three metal ions
Mg$^{+}$, C$^{\hbox{\tiny +3}}$, and O$^{\hbox{\tiny +5}}$ that are
responsible for the {\MgIIdblt}, {\CIVdblt}, and {\OVIdblt} doublets
commonly studied in absorption.  Over the $\log n\subH$--$T$ phase
space $-7 \leq \log n\subH \leq 0$ and $3 \leq \log T \leq 6$, the
logarithmic difference in the ionization corrections agreed with in
$\pm 0.05$.  This agreement is within typical uncertainties of
measured logarithmic column densities.

Future improvements to the code include (1) self shielding so that
optically thick cloud models can be treated, and (2) radiative
transfer through the simulation box to handle frequency dependent
shadowing from structures intervening to luminous sources.  These will
be reported in future papers.  

The stand-alone version of {\code}, is available on-line at {\it
  http://astronomy.nmsu.edu/cwc/Software/Ioncode/}.  This code is
useful for generating grids of optically thin model clouds as a
function of hydrogen density, temperature, and redshift.  The output
includes the equilibrium ionization fractions, number densities,
photoionization rates, and ionization and recombination rate
coefficients for all collisional processes for all ions.  Once
self-shielding is added, the code will be updated on-line and will be
capable of generating optically thick cloud models.




\acknowledgments

CWC, EK, and JRV were partially supported through grants HST-AR-12646
and HST-GO-13398 provided by NASA via the Space Telescope Science
Institute, which is operated by the Association of Universities for
Research in Astronomy, Inc., under NASA contract NAS 5-26555.  CWC
acknowledges assistance through the NASA New Mexico Space Grant
Consortium (NMSGC) Research Enhancement Program, which supported AM.
JRV acknowledges support through an NMSGC Graduate Research
Fellowship.  Thanks to R. Sutherland for helpful email exchanges and
to G. Ferland for insightful conversations during his visit to NMSU.
Much gratitude goes to D. Verner for providing publicly available
electronic tables for photoionization, recombination, collisional
ionization, and autoionization and some supporting computational
subroutines ({\it http://www.pa.uky.edu/~verner/}).  We are grateful
for the work of N. Badnell, R. Bingham, G. Duxbury, and H. Summers of
the Atomic and Molecular Diagnostic Processes in Plasmas group ({\it
  http://amdpp.phys.strath.ac.uk/tamoc/}), who provided radiative and
dielectronic recombination rates.  We also thank P. Stancil,
D. Schultz, J. Wang, M.  Rakovi\'c, J.  Kingdon, and A. Dalgarno, of
the Oakridge National Lab UGA Charge Transfer Database for
Astrophysics ({\it http://www-cfadc.phy.ornl.gov/astro/ps/data/}).

\end{document}